\pdfoutput=1

\documentclass[11pt,twoside,a4paper,cmspaper,final,collab]{cms-tdr}
\def\svnVersion{0f3b3dd-D}\def\svnDate{2020/12/14}\def\cmsCernNoTag{CERN-EP-2020-168}\def\cmsCernDate{\today}\def\cmsMessage{"Published in Physics Letters B as \href{http://dx.doi.org/10.1016/j.physletb.2020.136018}{\doi{10.1016/j.physletb.2020.136018}.}"}

\begin{document}\cmsNoteHeader{SMP-20-006}

\hyphenation{had-ron-i-za-tion}
\hyphenation{cal-or-i-me-ter}
\hyphenation{de-vices}
\newcommand{\mll}{\ensuremath{m_{\ell\ell}}\xspace}
\newcommand{\jet}{\ensuremath{\mathrm{j}}}
\newcommand{\mjj}{\ensuremath{m_{\jet\jet}}\xspace}
\newcommand{\detajj}{\ensuremath{\abs{\Delta\eta_{\jet\jet}}}\xspace}
\newcommand{\dphijj}{\ensuremath{\Delta\phi_{\jet\jet}}\xspace}
\newcommand{\tZq}{\ensuremath{\PQt\PZ\Pq}\xspace}
\newcommand{\WW}{\ensuremath{\PW^\pm\PW^\pm}\xspace}
\newcommand{\WWLL}{\ensuremath{\PW^\pm_{\mathrm{L}}\PW^\pm_{\mathrm{L}}}\xspace}
\newcommand{\WWT}{\ensuremath{\PW^\pm_{X}\PW^\pm_{\mathrm{T}}}\xspace}
\newcommand{\WWTT}{\ensuremath{\PW^\pm_{\mathrm{T}}\PW^\pm_{\mathrm{T}}}\xspace}
\newcommand{\WWLT}{\ensuremath{\PW^\pm_{\mathrm{L}}\PW^\pm_{\mathrm{T}}}\xspace}
\newcommand{\WWL}{\ensuremath{\PW^\pm_{\mathrm{L}}\PW^\pm_{X}}\xspace}
\newcommand{\WZ}{\ensuremath{\PW\PZ}\xspace}
\newcommand{\tVx}{\ensuremath{\PQt\PV\mathrm{x}}\xspace}
\newcommand{\PWT}{\ensuremath{\PW_{\mathrm{T}}}\xspace}
\renewcommand{\PWL}{\ensuremath{\PW_{\mathrm{L}}}\xspace}
\newlength\cmsTabSkip\setlength{\cmsTabSkip}{1ex}
\newlength\cmsFigWidth
\ifthenelse{\boolean{cms@external}}{\setlength\cmsFigWidth{0.49\textwidth}}{\setlength\cmsFigWidth{0.65\textwidth}} 

\cmsNoteHeader{SMP-20-006}

\title{Measurements of production cross sections of polarized same-sign $\PW$ boson pairs in association with two jets in proton-proton collisions at \texorpdfstring{$\sqrt{s} = 13\TeV$}{sqrt(s) = 13 TeV}}

\author*[inst1]{CMS experiment}

\date{\today}

\abstract{
  The first measurements of production cross sections of polarized same-sign $\WW$ boson pairs in proton-proton collisions are reported. The measurements are based on a data sample collected with the CMS detector at the LHC at a center-of-mass energy of $13\TeV$, corresponding to an integrated luminosity of $137\fbinv$. Events are selected by requiring exactly two same-sign leptons, electrons or muons, moderate missing transverse momentum, and two jets with a large rapidity separation and a large dijet mass to enhance the contribution of same-sign $\WW$ scattering events. An observed (expected) 95\% confidence level upper limit  of 1.17 (0.88)\unit{fb} is set on the production cross section for longitudinally polarized same-sign $\WW$ boson pairs. The electroweak production of same-sign $\WW$ boson pairs with at least one of the $\PW$ bosons longitudinally polarized is measured with an observed (expected) significance of 2.3 (3.1) standard deviations.}

\hypersetup{%
pdfauthor={CMS Collaboration},%
pdftitle={Measurements of production cross sections of polarized same-sign W boson pairs in association with two jets in proton-proton collisions at sqrts = 13 TeV},%
pdfsubject={CMS},%
pdfkeywords={CMS, diboson, electroweak, polarized, longitudinal}
}

\maketitle

\section{Introduction}
\label{sec:Introduction}

Vector boson scattering (VBS) processes probe the electroweak (EW) symmetry breaking mechanism at high energy scales. The unitarity of the tree-level amplitude of the scattering of longitudinally polarized gauge bosons at high energies is restored in the standard model (SM) by a Higgs boson with a mass lower than about 1$\TeV$~\cite{Lee:1977yc,Lee:1977eg}. The observation of a Higgs boson with a mass of about $125\GeV$~\cite{AtlasPaperCombination,CMSPaperCombination,CMSPaperCombination2} provides an explanation that $\PW$ and $\PZ$ gauge bosons acquire mass via the Brout--Englert--Higgs mechanism, but additional Higgs bosons may still play a role in the EW symmetry breaking. Modifications of the VBS cross section for the longitudinally polarized $\PW$ and $\PZ$ bosons are predicted in models of physics beyond the SM through modifications of the Higgs boson couplings to gauge bosons or through the presence of new resonances~\cite{Espriu:2012ih,Chang:2013aya}. The measurements of the longitudinally polarized scattering of the $\PW$ and $\PZ$ bosons provide complementary information to direct measurements of the Higgs boson couplings to gauge bosons~\cite{Khachatryan:2016vau, Sirunyan:2018egh}. Models of beyond SM physics that modify the cross sections of VBS processes with transversely polarized $\PW$ and $\PZ$ bosons are discussed in Ref.~\cite{Brass:2018hfw}.

\begin{figure*}[htbp]
\centering
\includegraphics[width=0.30\textwidth]{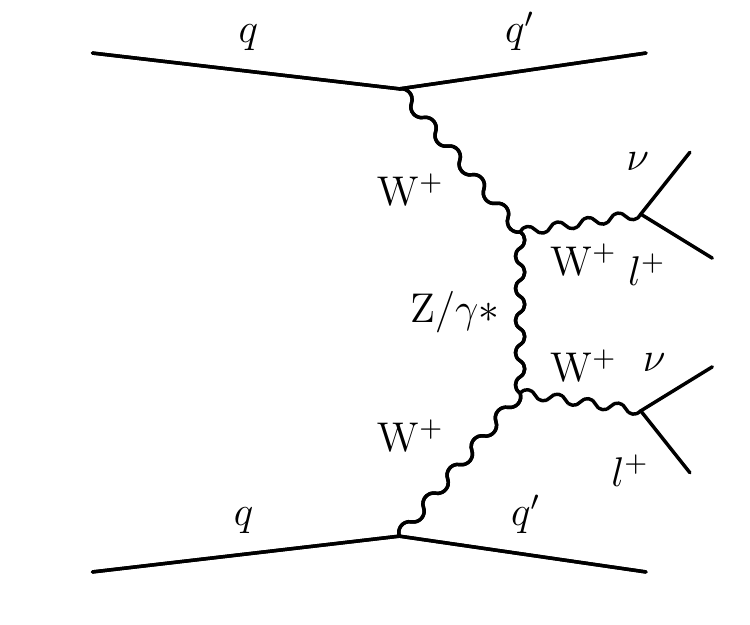}
\includegraphics[width=0.30\textwidth]{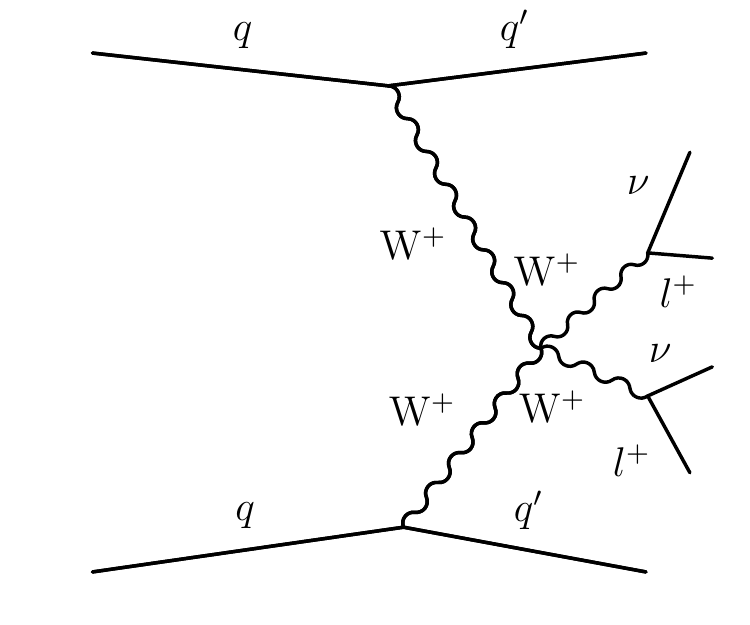}
\includegraphics[width=0.30\textwidth]{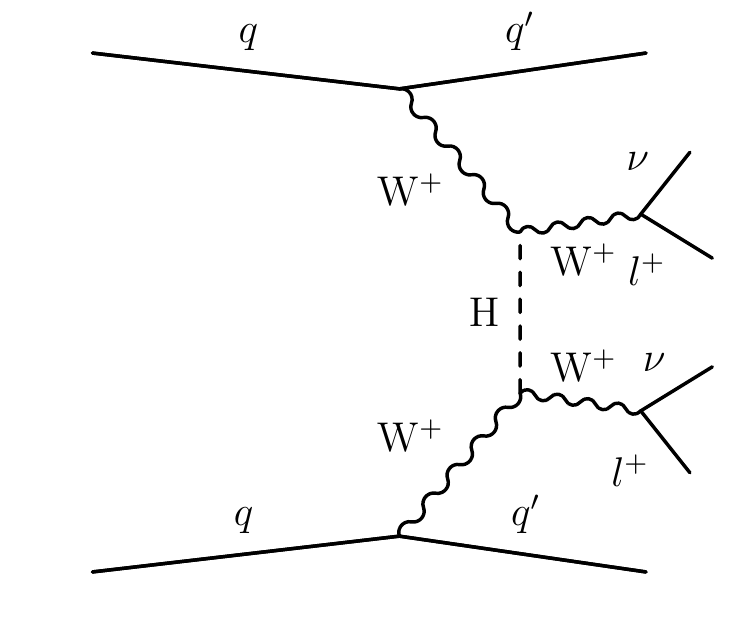}
\caption{
  Illustrative Feynman diagrams of VBS processes, where $\PW$ bosons are radiated from incoming quarks (q), contributing to the EW-induced production of events containing two forward jets and $\WW$ boson pairs decaying to leptons. Diagrams with the triple gauge coupling vertex (left), the quartic gauge coupling vertex (center), and the $t$-channel Higgs boson exchange (right) are shown.}
\label{fig:feynman}
\end{figure*}

At the CERN LHC, VBS interactions are characterized by the presence of two gauge bosons in association with two forward jets that have a large rapidity separation. They are part of a class of processes contributing to the same-sign $\WW$ production in association with two jets that proceeds via the EW interaction at tree level, $\mathcal{O}(\alpha^4)$, where $\alpha$ is the EW coupling, referred to as EW $\WW$ production. The leptonic decay mode $\WW \to \ell^\pm\PGn\ell'^\pm\PGn$, where both $\PW$ bosons decay into electrons or muons, $\ell, \ell' = \Pe$, $\Pgm$, is a promising final state to study the polarized scattering from gauge bosons. The background contribution of the quantum chromodynamics (QCD) induced production of $\PW^\pm\PW^\pm$ boson pairs in association with two jets with tree-level contributions at $\mathcal{O}(\alpha^2\alpS^2)$, where $\alpS$ is the strong coupling, is small. Figure~\ref{fig:feynman} shows representative Feynman diagrams of VBS processes involving self-interactions between gauge bosons through triple and quartic gauge couplings and the $t$-channel Higgs boson exchange.

The unpolarized EW $\WW$ production has been previously measured at the LHC in the leptonic decay modes at $\sqrt{s}=8$ and $13\TeV$~\cite{Aad:2014zda,Khachatryan:2014sta,Sirunyan:2017ret,Aaboud:2019nmv,Sirunyan:2020gyx}. The first differential cross section measurements were reported in Ref.~\cite{Sirunyan:2020gyx}. This Letter presents the first measurement of the EW production cross sections for polarized same-sign $\WW$ boson pairs. The data sample of proton-proton ($\Pp\Pp$) collisions at $\sqrt{s}=13\TeV$ corresponds to an integrated luminosity of $137\fbinv$~\cite{CMS-PAS-LUM-17-001,CMS-PAS-LUM-17-004,CMS-PAS-LUM-18-002}, collected with the CMS detector~\cite{Chatrchyan:2008aa} in three LHC operating periods during the years 2016, 2017, and 2018. The three data sets are analyzed independently, with appropriate calibrations and corrections, because of the various LHC operating conditions and the upgrades in the performance of the CMS detector. Candidate events contain exactly two identified same-sign leptons, moderate missing transverse momentum, and two jets with a large rapidity separation and a high dijet mass. 

In the $\WW$ channel, each of the \PW bosons can be polarized either longitudinally (\PWL) or transversely (\PWT), leading to three distinct contributions $\WWLL$, $\WWLT$, and $\WWTT$. Ideally, we would measure all three contributions separately, but the current data sample size is too limited. Therefore, two maximum-likelihood fits are performed: one for $\WWLL$ and $\WWT$; and another for $\WWL$ and  $\WWTT$. The index $X$ indicates either of the two polarization states. The event kinematical properties are used to extract the various contributions. Two sets of results are reported with the helicity eigenstates defined either in the $\PW\PW$ center-of-mass reference frame or in the initial-state parton-parton one.

\section{The CMS detector}
\label{sec:cms}

The central feature of the CMS apparatus is a superconducting solenoid of 6\unit{m} internal diameter, providing a magnetic field of 3.8\unit{T}. Within the solenoid volume are a silicon pixel and strip tracker, a lead tungstate 
crystal electromagnetic calorimeter (ECAL), and a brass and scintillator hadron calorimeter, each composed of a barrel and two endcap sections. Forward calorimeters extend the pseudorapidity ($\eta$) coverage provided by the barrel and endcap detectors. Muons are detected in gas-ionization detectors embedded in the steel flux-return yoke outside the solenoid. A more detailed description of the CMS detector, together with a definition of the coordinate system used and the relevant kinematic variables, is given in Ref.~\cite{Chatrchyan:2008aa}.

The first level of the CMS trigger system, composed of custom hardware processors, uses information from the calorimeters and muon detectors to select events of interest with a latency of less than 4\mus. 
The second level, known as the high-level trigger, consists of a farm of processors running a version of the full event reconstruction software optimized for fast processing, and reduces the event rate to about 1\unit{kHz} before data storage~\cite{Khachatryan:2016bia}.

\section{Signal and background simulation}
\label{sec:samples}

Several Monte Carlo (MC) event generators are used to simulate the signal and background contributions. Three independent sets of simulated events for each process are needed to match the data-taking conditions in the various years. All generated events are processed through a simulation of the CMS detector based on \GEANTfour~\cite{Geant} and are reconstructed with the same algorithms used for data. Additional $\Pp\Pp$ interactions in the same and nearby bunch crossings, referred to as pileup, are also simulated. The distribution of the number of pileup interactions in the simulation is adjusted to match the one observed in the data. The average number of pileup interactions was 23 (32) in 2016 (2017 and 2018).

The SM EW $\WWLL$, $\WWLT$, and $\WWTT$ signal processes, where both bosons decay leptonically, are separately simulated using $\MGvATNLO$ 2.7.2, with the implementation of polarized parton scattering~\cite{Frederix2012,Alwall:2014hca,BuarqueFranzosi:2019boy}, at leading order (LO) with six EW ($\mathcal{O}(\alpha^6)$) and zero QCD vertices. The NNPDF 3.1 next-to-next-to-leading-order (NNLO)~\cite{Ball:2017nwa} parton distribution functions (PDFs) are used. Signal processes are simulated with the helicity eigenstates defined either in the $\WW$ center-of-mass reference frame or in the initial parton-parton reference frame. The \textsc{Phantom} 1.5.1 generator~\cite{Ballestrero:2007xq,Ballestrero:2011pe} uses the on-shell projection technique for the predictions of the signal processes as discussed in Ref.~\cite{Ballestrero:2017bxn}. The $\MGvATNLO$ predictions show satisfactory agreement within the statistical uncertainties with the \textsc{Phantom} predictions in the relevant fiducial region, defined in Section~\ref{sec:results}, for this analysis. Comparisons of the $\MGvATNLO$ predictions with predictions based on the on-shell projection technique are reported in Ref.~\cite{BuarqueFranzosi:2019boy}. The small contributions of off-shell and nonresonant production~\cite{Ballestrero:2017bxn} are not included in the simulated signal samples and amount to 1--2\% in the fiducial region.

The full next-to-leading-order (NLO) QCD and EW corrections for the leptonic unpolarized $\WW$ scattering process have been computed~\cite{Biedermann:2016yds,Biedermann:2017bss}, and they reduce the LO cross section for the EW $\WW$ process by approximately 10--15\%, with the correction increasing in magnitude to up to 25\% with increasing dilepton and 
dijet masses. The NLO corrections for the $\WWLL$, $\WWLT$, and $\WWTT$ processes are not known. The corrections for the unpolarized EW $\WW$ process at orders of $\mathcal{O}(\alpS\alpha^6)$ and 
$\mathcal{O}(\alpha^7)$ are applied to the $\MGvATNLO$ LO cross sections for the $\WWTT$ process. Only the corrections at order of $\mathcal{O}(\alpS\alpha^6)$ are applied to the $\MGvATNLO$ LO cross sections for the $\WWLL$ and $\WWLT$ processes because the corrections at order $\mathcal{O}(\alpha^7)$ are expected to be smaller for the $\WWLL$ and $\WWLT$ processes compared to the size of the corresponding corrections for the unpolarized EW $\WW$ process~\cite{Denner:2000jv}. There is a negligible effect in the measured cross sections from differences in the event kinematical properties caused by the treatment of the NLO corrections. 

The EW $\WZ$ background process is simulated with $\MGvATNLO$~2.4.2 at order $\mathcal{O}(\alpha^6)$. The 
QCD-induced $\WZ$ process is simulated at LO with up to three additional partons in the matrix element calculations using the $\MGvATNLO$ generator with at least one QCD vertex at tree level. The different jet multiplicities 
are merged using the MLM scheme~\cite{MLMmerging} to match matrix element and parton shower jets. The $\MGvATNLO$ generator is also used to simulate the QCD-induced $\WW$ process. 

The interference between the EW and QCD diagrams for the $\WW$ and $\WZ$ processes is generated with $\MGvATNLO$ including the contributions of order $\alpS\alpha^5$. The relative contributions in the fiducial region of the 
interference term between the EW and the QCD diagrams for the $\WWLL$, $\WWLT$, and $\WWTT$ processes are comparable to the relative contributions of the $\WWLL$, $\WWLT$, and $\WWTT$ processes to the EW $\WW$ cross section. The interferences between the signal processes are expected  to be small~\cite{BuarqueFranzosi:2019boy}, and good agreement is observed between the incoherent sum of the polarized cross sections and the unpolarized cross sections for the distributions of the observables.

The $\POWHEG$~v2~\cite{Frixione:2002ik,Nason:2004rx,Frixione:2007vw,Alioli:2008gx,Alioli:2010xd} generator is used to simulate the $\ttbar$, $\PQt\PW$, and other diboson processes at NLO accuracy in QCD. Production of $\ttbar\PW$, $\ttbar\PZ$, $\ttbar\gamma$, and triple vector boson ($\PV\PV\PV$) events is simulated at 
NLO accuracy in QCD using the $\MGvATNLO$~2.2.2 (2.4.2) generator for 2016 (2017 and 2018)~\cite{Frederix2012,Alwall:2014hca} samples. The $\tZq$ process is simulated at NLO in the four-flavor scheme using $\MGvATNLO$~2.3.3. The $\tZq$ MC simulation is normalized using a cross section computed at NLO with $\MGvATNLO$ in the five-flavor scheme, following the procedure described in Ref.~\cite{Sirunyan:2017nbr}. 
The double parton scattering $\WW$ production is generated at LO using $\PYTHIA$ 8.226 (8.230)~\cite{Sjostrand:2014zea} for 2016 (2017 and 2018) samples.

The NNPDF~3.0 NLO \cite{Ball:2014uwa} (NNPDF~3.1 NNLO~\cite{Ball:2017nwa}) PDFs are used for generating all 2016 (2017 and 2018) background samples. For all processes, the parton showering and hadronization are simulated using $\PYTHIA$~8.226 (8.230) for 2016 (2017 and 2018). The modeling of the underlying event is done using the CUETP8M1~\cite{Skands:2014pea,Khachatryan:2015pea} (CP5~\cite{Sirunyan:2019dfx}) tune for simulated samples corresponding to the 2016 (2017 and 2018) data.

\section{Event reconstruction and selection}
\label{sec:objects}

Events are reconstructed using the CMS particle-flow (PF) algorithm~\cite{Sirunyan:2017ulk} that reconstructs and identifies each individual particle with an optimized combination of all subdetector 
information. The missing transverse momentum vector $\ptvecmiss$ is defined as the projection onto the plane perpendicular to the beam axis of the negative vector sum of the momenta of all reconstructed PF objects in an event. Its magnitude is referred to as $\ptmiss$. 

Jets are reconstructed by clustering PF candidates using the anti-\kt algorithm~\cite{Cacciari:2008gp,Cacciari:2011ma} with a distance parameter of 0.4. Jets are calibrated in the simulation, and separately in data, accounting for energy deposits of neutral particles from pileup and any nonlinear detector response~\cite{Khachatryan:2016kdb}. The effect of pileup is mitigated through a charged-hadron subtraction technique, which removes the energy of charged hadrons not originating from the primary vertex (PV)~\cite{Sirunyan:2020foa} of the event. Corrections to jet energies to account for the detector response are propagated to $\ptmiss$~\cite{Sirunyan:2019kia}. The PV is defined as the vertex with the largest value of summed physics-object $\pt^2$. The physics objects are derived from only the tracks assigned to the vertex as inputs by clustering them into jets, including leptons. The $\ptmiss$ is also recalculated only from those jets by summing their negative $\pt$ vectors.

Electrons and muons are reconstructed by associating a track reconstructed in the tracking detectors with either a cluster of energy in the ECAL~\cite{Khachatryan:2015hwa, CMS-DP-2018-017} or a track in the muon system~\cite{Sirunyan:2018fpa}. Electron and muon candidates must pass certain identification criteria to be further selected in the analysis. For the ``loose'' identification, they must satisfy $\pt > 10\GeV$ and $\abs{\eta} < 2.5$ (2.4) for electrons (muons). At the final stage of the lepton selection the ``tight'' working points criteria following the definitions provided in Refs.~\cite{Khachatryan:2015hwa,Sirunyan:2018fpa} are chosen, including requirements on the impact parameter of the candidates with respect to the PV and their isolation with respect to other particles in the event~\cite{Sirunyan:2018egh}. 

For electrons, the background contribution coming from a mismeasurement of the track charge is not negligible. The sign of this charge is evaluated using three observables that measure the electron curvature applying different methods; requiring all three charge evaluations to agree reduces this background contribution by a factor of five with an efficiency of about 97\%~\cite{Khachatryan:2015hwa}. The charge mismeasurement is negligible for muons~\cite{Chatrchyan:2009ae,Sirunyan:2019yvv}.  

Collision events are collected using single-electron (single-muon) triggers that require the presence of an isolated lepton with $\pt>27$ (24)$\GeV$. In addition, a set of dilepton triggers with lower $\pt$ thresholds, with a threshold of $23\GeV$ or lower for the leading lepton and with a threshold of $8\GeV$ for the subleading lepton, is used. This ensures a trigger efficiency above 99\% for events that satisfy the subsequent offline selection.

Several selection requirements are used to isolate the $\WW$ topology defining the signal region (SR) while reducing the contributions of background processes. Candidate events contain exactly two isolated same-sign charged leptons and at least two jets with $\pt^{\mathrm{j}}>50\GeV$ and $\abs{\eta}<4.7$. Jets that are within $\Delta R = \sqrt{\smash[b]{(\Delta \phi)^{2} + (\Delta \eta)^{2}}} < 0.4$ of one of the identified leptons are not used in the analysis. Here $\Delta\phi$ and $\Delta\eta$ refer to the differences in the azimuthal angle $\phi$ and $\eta$ of the jet and the charged-lepton candidate, respectively. Because of the presence of undetected neutrinos in the signal events, $\ptmiss$ is required to exceed $30\GeV$. 

The $\WW$ SR selection requires one of the same-sign leptons to satisfy $\pt>25\GeV$ and the other $\pt>20\GeV$. The mass of the dilepton pair $\mll$ must be greater than $20\GeV$. Candidate 
events in the dielectron final state within $15\GeV$ of the nominal $\PZ$ boson mass $m_{\PZ}$~\cite{10.1093/ptep/ptaa104} are rejected to reduce the number of $\PZ$~boson background events where the charge of one of the electron candidates is misidentified. 

The VBS topology is targeted by requiring the two highest-$\pt$ jets to have a dijet mass $\mjj>500\GeV$ and a pseudorapidity separation $\detajj>2.5$. The $\PW$ bosons in the VBS topologies are mostly produced in the central rapidity region with respect to the two selected jets. The candidate $\WW$ events are required to satisfy 
$\max(\mathrm{z}_{\ell}^{*})<0.75$, where $\mathrm{z}_{\ell}^{*}=\abs{\eta^{\ell} - (\eta^{\jet_{1}} + \eta^{\jet_{2}})/2}/\detajj$ is the Zeppenfeld variable~\cite{Rainwater:1996ud}, $\eta^{\ell}$ is the pseudorapidity of one of the selected leptons, and $\eta^{\jet_{1}}$ and $\eta^{\jet_{2}}$ are the pseudorapidities of the two candidate VBS jets.

Candidate events with one or more jets with $\pt>20\GeV$ and $\abs{\eta}<2.4$ that are consistent with the fragmentation of a bottom quark are rejected to reduce the number of top quark background events. The \textsc{DeepCSV} \PQb\ tagging algorithm~\cite{Sirunyan:2017ezt} is used for this selection. For the chosen working point, the efficiency to select \PQb\ quark jets is about 70\% and the rate for incorrectly tagging jets originating from the hadronization of gluons or $\cPqu$, $\cPqd$, $\cPqs$ quarks is about 1\%. The rate for incorrectly tagging jets originating from the hadronization of $\cPqc$ quarks is about 10\%. The selection requirements to define the same-sign $\WW$ SR are summarized in Table~\ref{tab:selectioncutsSR}. 

\begin{table*}[htb]
\centering
\topcaption{
  Summary of the requirements defining the $\WW$ SR. The 
  $\abs{\mll - m_{\PZ}}$ requirement is applied to the dielectron final state 
  only.\label{tab:selectioncutsSR}}
\begin{tabular} {lc}
\hline
  Variable & Requirement \\
\hline
Leptons                               & Exactly 2 same-sign leptons, $\pt>25/20\GeV$ \\
$\pt^{\mathrm{j}}$                    & $>$50\GeV                   \\
$\abs{\mll - m_{\PZ}}$               & $>$15\GeV ($\Pe\Pe$)        \\
$\mll$                                & $>$20\GeV                   \\
$\ptmiss$                             & $>$30\GeV                   \\
\PQb\ quark veto                      & Required                    \\
$\mathrm{Max}(\mathrm{z}_{\ell}^{*})$ & $<$0.75                    \\
$\mjj$                                & $>$500\GeV                 \\
$\detajj$                             & $>$2.5                     \\
\hline
\end{tabular}
\end{table*}

\section{Extracting polarization information}
\label{sec:polarization}

In the $\WW$ channel, the $\PW$ bosons can each be either longitudinally or transversely polarized leading to different kinematic distributions, reflected in the kinematical properties of the two leptons, the two jets, and $\ptvecmiss$. The $\PWL$ bosons tend to be radiated at a smaller angle with respect to the incoming quark direction, resulting in a smaller $\PWL$ boson $\pt$ compared to the radiated $\PWT$ boson $\pt$. In addition, there are differences in the behavior of the scattering amplitudes as a function of the $\WW$ center-of-mass energy and the scattering angle~\cite{Doroba:2012pd}. 

\begin{figure*}[htbp]
\centering
\includegraphics[width=0.49\textwidth]{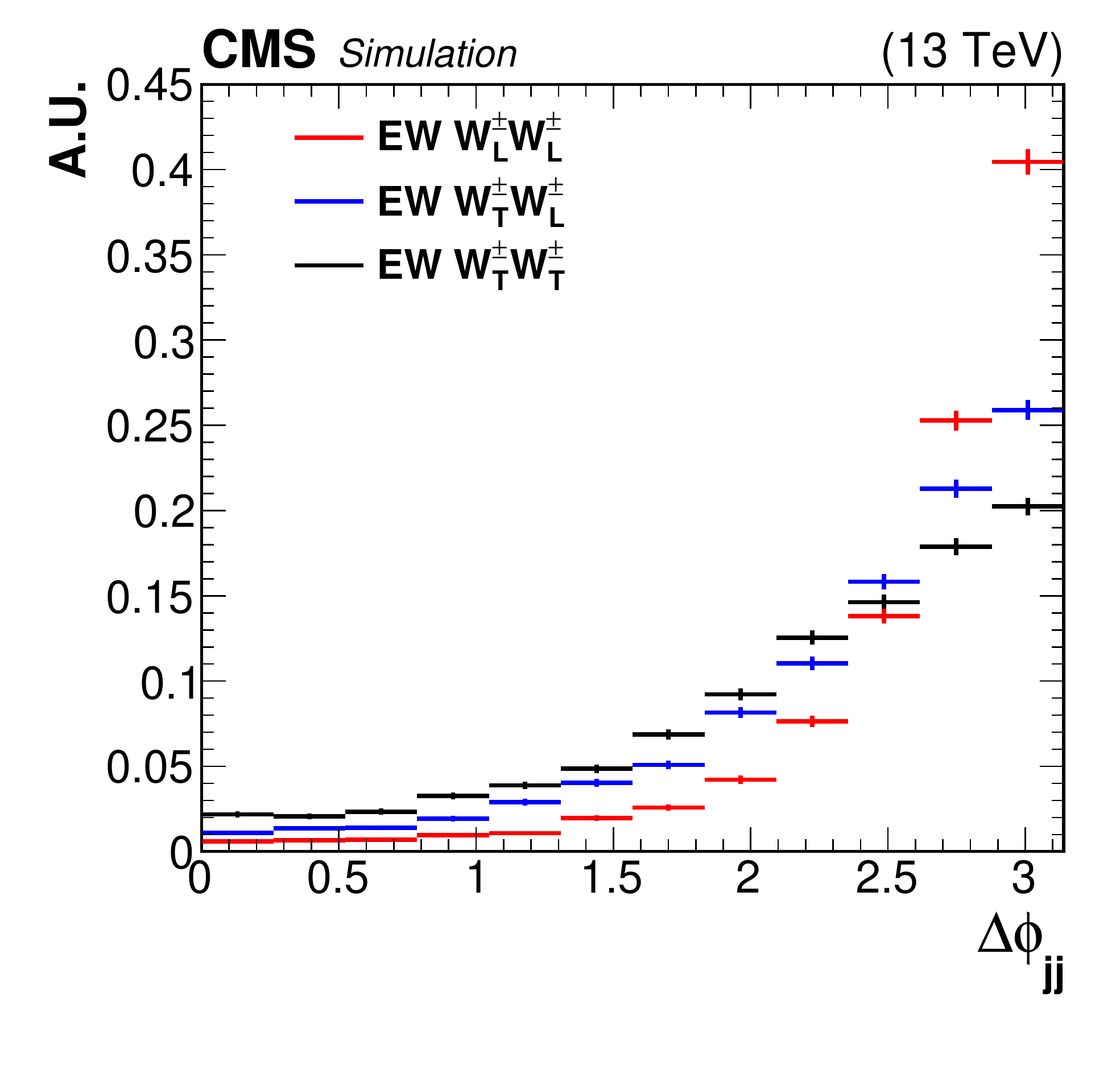}
\includegraphics[width=0.49\textwidth]{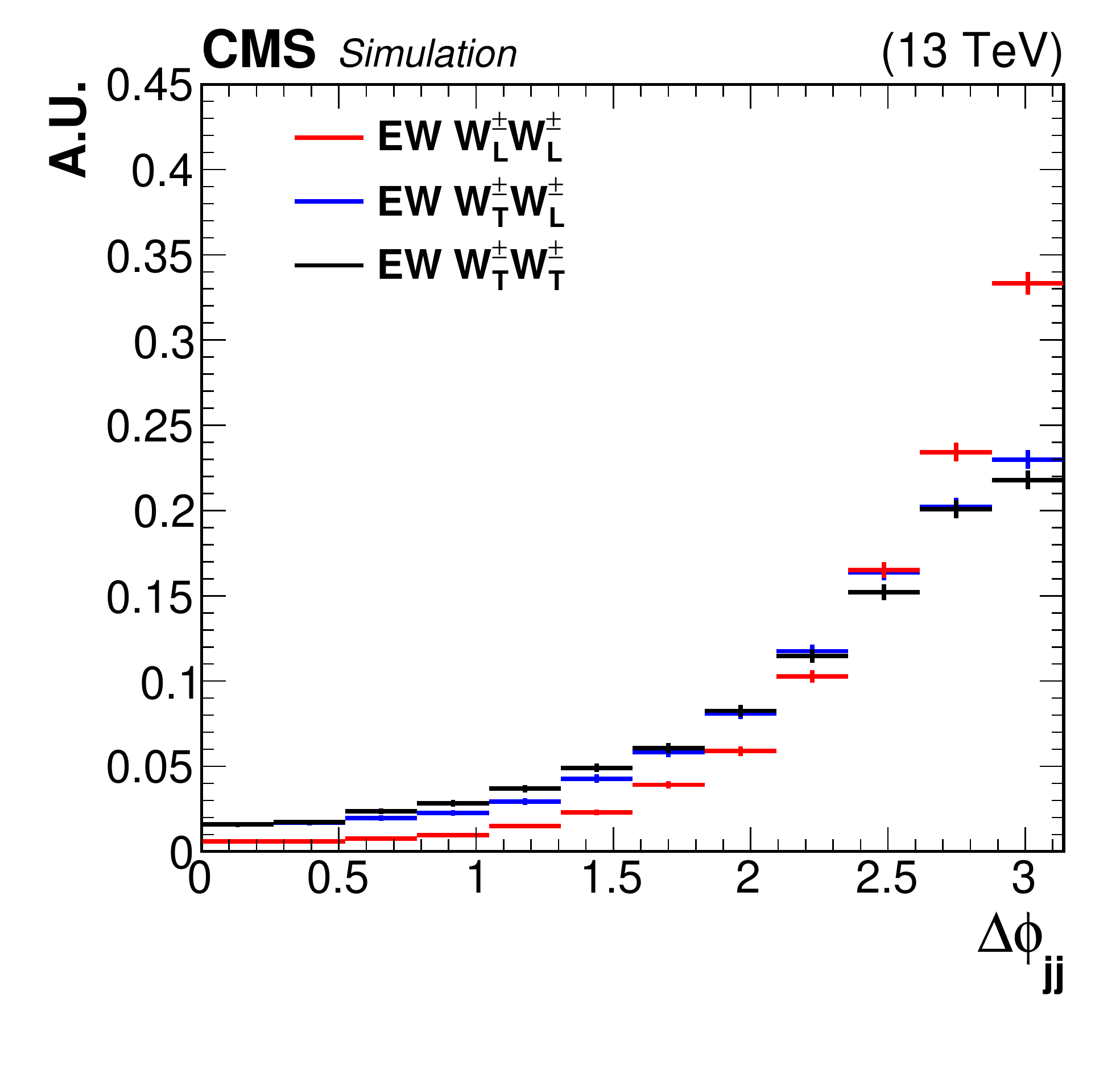}
\includegraphics[width=0.49\textwidth]{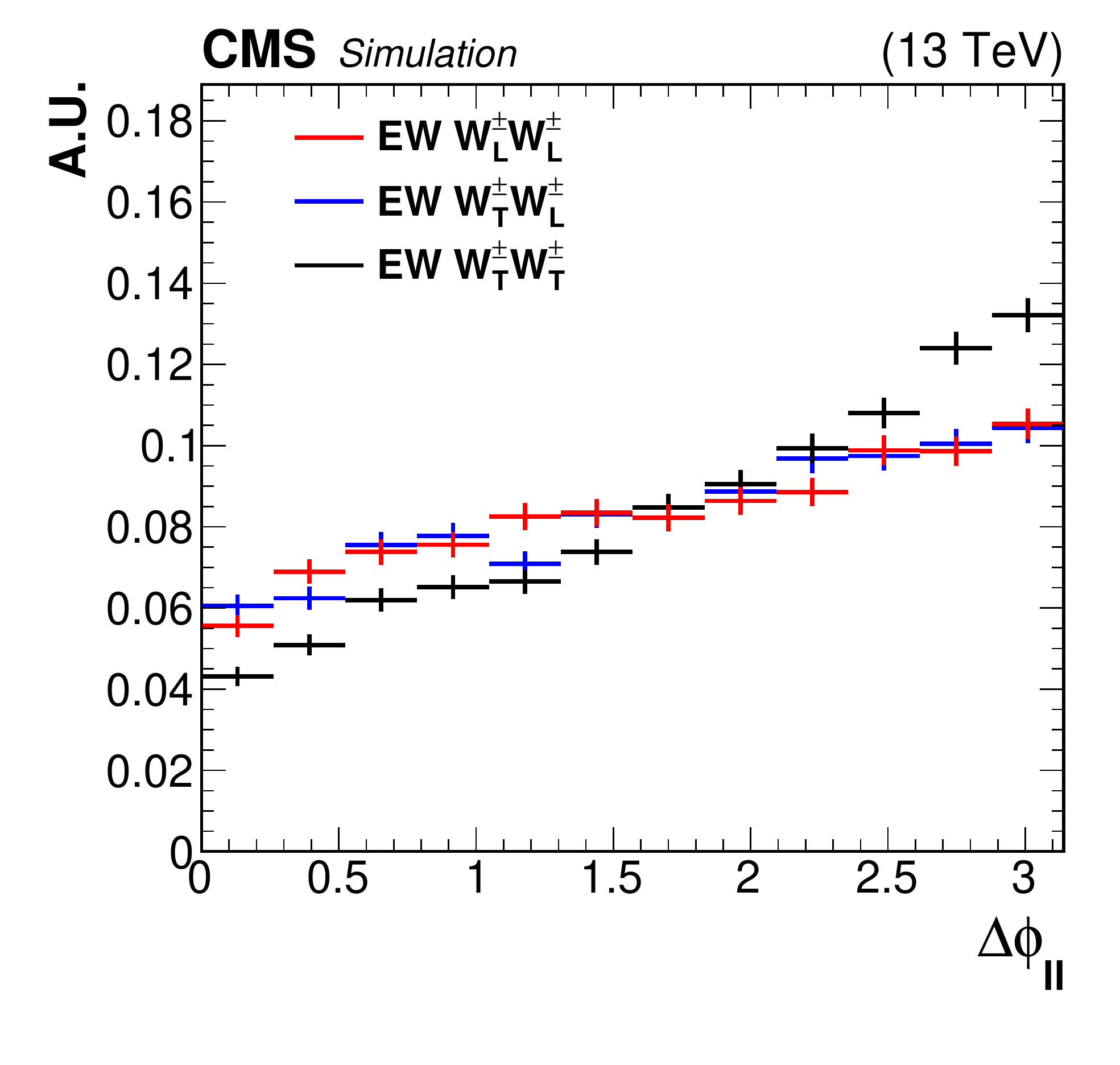}
\includegraphics[width=0.49\textwidth]{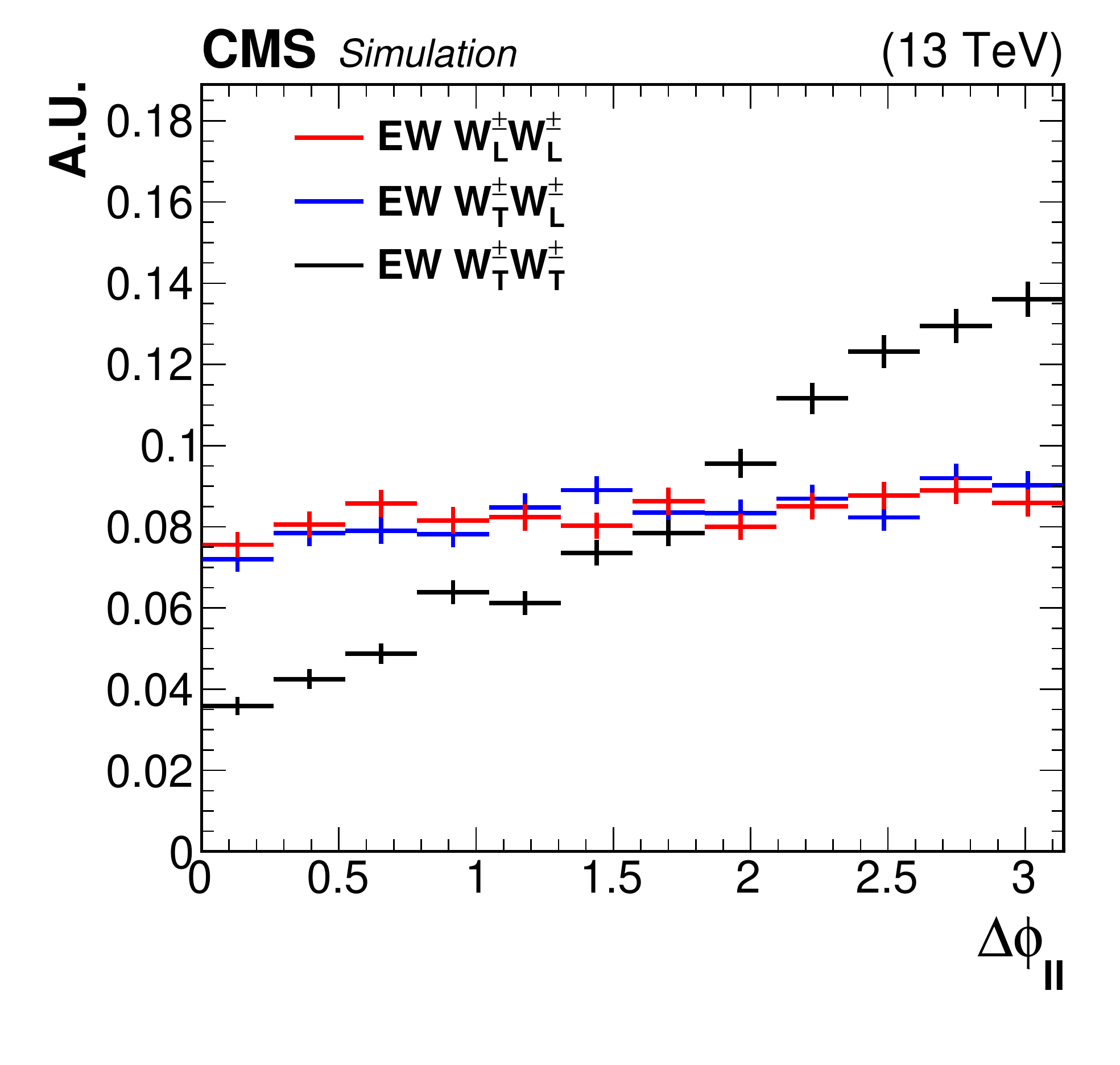}
\includegraphics[width=0.49\textwidth]{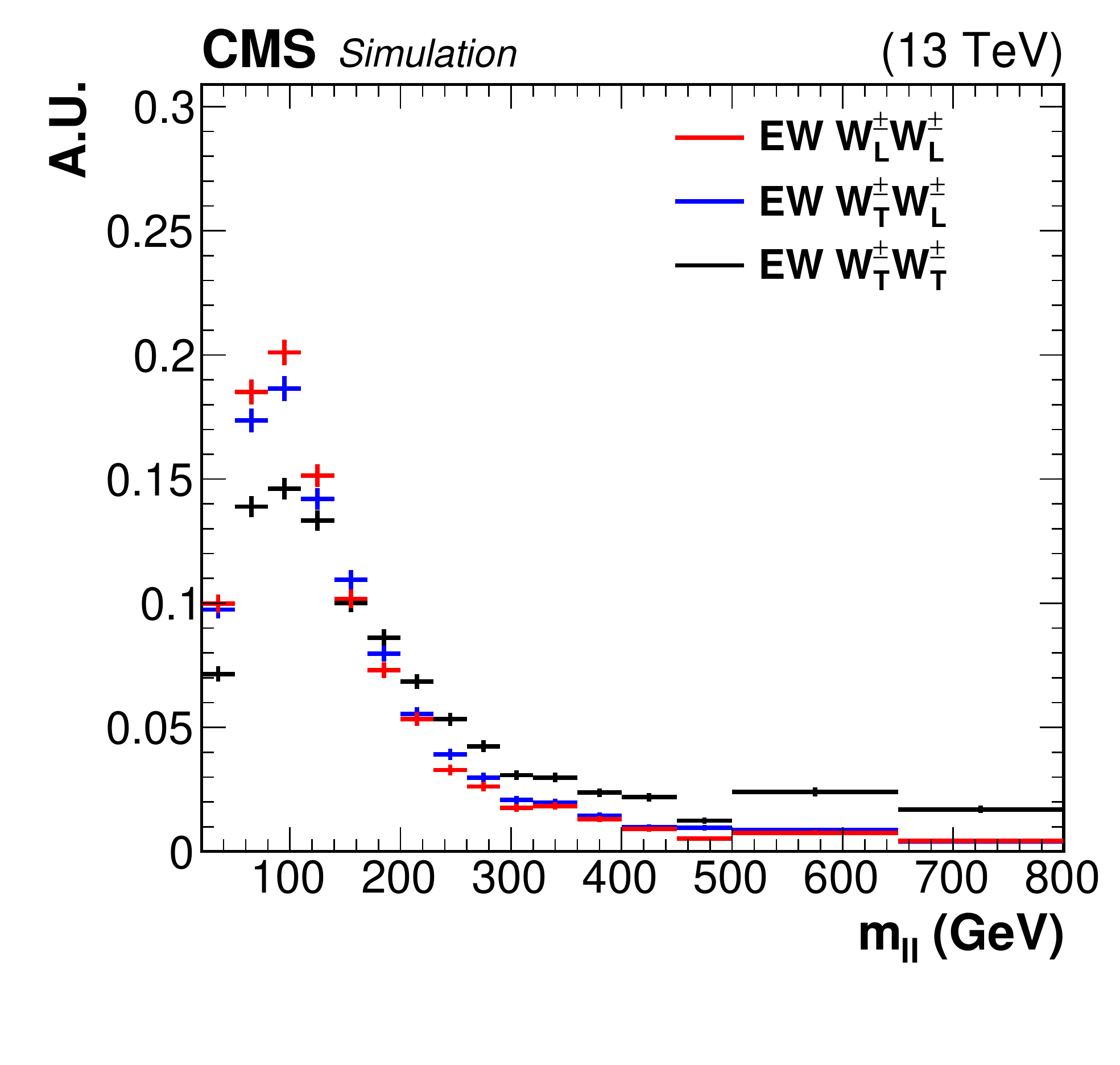}
\includegraphics[width=0.49\textwidth]{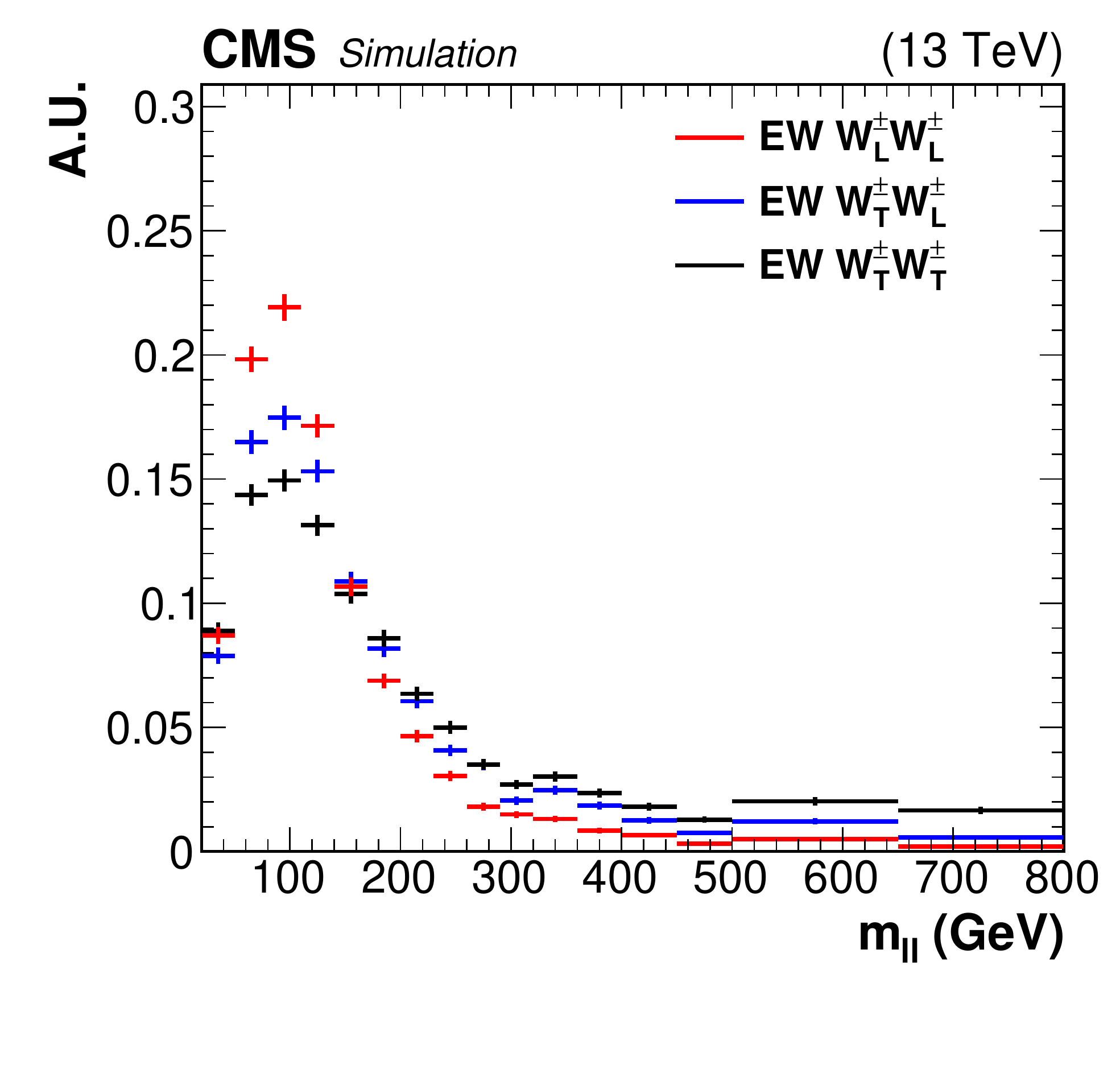}
\centering
\caption{Generator level distributions of $\dphijj$ (upper), $\Delta\phi_{\ell\ell}$ (center), and $\mll$ (lower) in the fiducial region for the $\WWLL$, $\WWLT$, and $\WWTT$ processes with the helicity eigenstates defined in the parton-parton (left) and $\WW$ (right) center-of-mass reference frames. The distributions are normalized to unit area. The error bars represent the uncertainties associated with the limited numbers of simulated events.}
\label{fig:sswwlong_genshape}
\end{figure*}

Multivariate techniques are used to enhance the separation between the different processes. We implement boosted decision trees (BDTs) with gradient boosting using the \textsc{tmva} package~\cite{Hocker:2007ht}. Two different BDTs, referred to as the signal BDTs, are trained on simulated events to separate either the $\WWLL$ and $\WWT$ processes or the $\WWL$ and $\WWTT$ processes. Several discriminating observables are used as the inputs to the BDTs, including the jet and lepton kinematical properties and $\ptmiss$, as summarized in Table~\ref{table:BDT_variables}. The distributions of these observables are taken from the SM predictions. Hypothetical modifications due to beyond the SM physics are assumed to impact only the production rates, but not the kinematic distributions of sensitive variables. Angular variables are included, such as the difference in the azimuthal angles between the leading and subleading jets ($\dphijj$) and leptons ($\Delta\phi_{\ell\ell}$), and the $\Delta R$ between the leading (subleading) jet and the dilepton system $\Delta R_{\jet1,\ell\ell}$ ($\Delta R_{\jet2,\ell\ell}$ ). The dilepton $\pt^{\ell\ell}$, $\mll$, and the transverse diboson mass $\mT^{\PW\PW}$ as defined in Ref.~\cite{Sirunyan:2020gyx} are also considered. The kinematic variable $(\pt^{\ell_1}\pt^{\ell_2})/(\pt^{\jet1}\pt^{\jet2})$ proposed in Ref.~\cite{Doroba:2012pd} is also included in the BDT inputs. A larger set of discriminating observables was studied, but only variables that improve the sensitivity and show some separation are retained. The distributions of $\dphijj$ (upper), $\Delta\phi_{\ell\ell}$ (middle), and $\mll$ (lower) at the generator level for the $\WWLL$, $\WWLT$, and $\WWTT$ processes with the helicity eigenstates defined in the parton-parton (left) and $\WW$ (right) center-of-mass reference frames are shown in Fig.~\ref{fig:sswwlong_genshape}. The signal extraction was also compared with a deep neural network using the \textsc{Keras}~\cite{chollet2015keras} deep learning library, interfaced with the \textsc{TensorFlow}~\cite{tensorflow2015-whitepaper} library, which led to a consistently good performance.

\begin{table*}[htb]
\centering
\topcaption{\label{table:BDT_variables}
  List and description of all the input variables for the signal BDT trainings.}
\renewcommand{\arraystretch}{1.5}
\begin{tabular}{lc}
Variables & Definitions\\
\hline
$\dphijj$	         	                       & Difference in azimuthal angle between the leading and subleading jets \\
$\pt^{\jet1}$		         	               & $\pt$ of the leading jet \\
$\pt^{\jet2}$		         	               & $\pt$ of the subleading jet \\[\cmsTabSkip]

$\pt^{\ell_1}$                   	               & Leading lepton $\pt$ \\
$\pt^{\ell_2}$                   	               & Subleading lepton $\pt$ \\
$\Delta\phi_{\ell\ell}$          	               & Difference in azimuthal angle between the two leptons \\
$\mll$                           	               & Dilepton mass \\
$\pt^{\ell\ell}$                 	               & Dilepton $\pt$ \\
$\mT^{\PW\PW}$                   	               & Transverse $\PW\PW$ diboson mass \\[\cmsTabSkip]

$z_{\ell_1}^{*}$	         	               & Zeppenfeld variable of the leading lepton \\
$z_{\ell_2}^{*}$	         	               & Zeppenfeld variable of the subleading lepton \\
$\Delta R_{\jet1,\ell\ell}$             	       & $\Delta R$ between the leading jet and the dilepton system \\
$\Delta R_{\jet2,\ell\ell}$             	       & $\Delta R$ between the subleading jet and the dilepton system \\
$(\pt^{\ell_1}\pt^{\ell_2})/(\pt^{\jet1}\pt^{\jet2})$  & Ratio of $\pt$ products between leptons and jets \\
$\ptmiss$                        	               & Missing transverse momentum \\
\hline
\end{tabular}
\end{table*}

\section{Background estimation}
\label{sec:backgrounds}

A combination of methods based on control samples in data and simulation is used to estimate background contributions. Uncertainties related to the theoretical and experimental predictions are described in Section~\ref{sec:systematics}. The electron charge misidentification in simulation is corrected to reproduce the rate measured in data. Using $\PZ \to \Pe\Pe$ events, the misidentification rate is about 0.01\% 
(0.3\%) in the barrel (endcap) region~\cite{Khachatryan:2015hwa}. Oppositely charged dilepton final states from $\ttbar$, $\PQt\PW$, $\PW^{+}\PW^{-}$, and Drell--Yan processes contribute to the background from charge misidentification. 

The nonprompt lepton backgrounds originating from leptonic decays of heavy quarks, hadrons misidentified as leptons, and electrons from photon conversion are suppressed by the identification and isolation requirements 
imposed on electrons and muons. The remaining contribution from the nonprompt lepton background is estimated directly from data following the technique described in Ref.~\cite{Khachatryan:2014sta}, where the yield in a sample of data events dominated by jet production is extrapolated to the signal region using efficiencies for loosely identified leptons to pass the standard lepton selection criteria. A normalization uncertainty of 20\% is assigned for the nonprompt lepton background to include possible differences in the composition of jets between the data sample used to derive these efficiencies and the data sample in the $\WW$ SR~\cite{Sirunyan:2018egh}.

Several background-enriched control regions (CRs), disjoint from one another and from the SR, are used to select event samples enriched with $\WZ$, nonprompt lepton, $\tZq$, and $\PZ\PZ$ background events. The $\WZ$ CR is defined by requiring three leptons where the opposite-sign same-flavor leptons from the $\PZ$ boson candidate have $\pt>25$ and $10\GeV$ with the dilepton mass within $15\GeV$ of the nominal $\PZ$ boson mass. In events with three same-flavor leptons, the opposite-sign lepton pair with the dilepton mass closest to $m_{\PZ}$ is associated with the $\PZ$ boson. The remaining lepton with $\pt>20\GeV$ is associated with the $\PW$ boson. In addition, the trilepton mass $m_{\ell\ell\ell}$ is required to be greater than $100\GeV$ and $\mathrm{max}(\mathrm{z}_{\ell}^{*})$ must be less than 1.0. Distributions of several kinematic variables in the $\WZ$ CR are reported in Ref.~\cite{Sirunyan:2020gyx}.

The nonprompt lepton CR is defined by requiring the same selection as for the $\WW$ SR, but with the \PQb\ jet veto requirement inverted. The selected sample is enriched with events from the nonprompt lepton background and 
dominated by semileptonic $\ttbar$ events. Similarly, the $\tZq$ CR is defined by requiring the same selection as the $\WZ$ CR, but with the \PQb\ quark veto requirement inverted. The selected sample is dominated by the $\tZq$ background process. Finally, the $\PZ\PZ$ CR requirements select events with four leptons with the same VBS requirements as the $\WW$ SR. The four CRs are used to estimate the normalization of the main background processes from data. All other background processes are estimated from simulation after applying corrections to account for small differences between data and simulation as detailed in Section~\ref{sec:systematics}.

To distinguish EW $\WW$ production from the SM background processes before extracting the individual polarizations, a BDT is trained using the TMVA package~\cite{Hocker:2007ht}. Several discriminating observables listed in Table~\ref{table:BDT_variables2} are used as inputs to this BDT, which we will refer to as the inclusive 
BDT. The values of $\mjj$ and $\detajj$ are powerful because VBS topologies typically have large values for the dijet mass and pseudorapidity separation~\cite{Sirunyan:2020gyx}.  The SM background processes are dominated by the nonprompt lepton background contribution, which comes mainly from top quark production. A large training background sample of simulated events is obtained by using oppositely charged dilepton events from top quark production.

\begin{table*}[htb]
  \centering
 {
\topcaption{\label{table:BDT_variables2} List and description of the input variables for the inclusive BDT training.}
\renewcommand{\arraystretch}{1.5}
\begin{tabular}{ l c c}
Variables & Definitions\\
\hline
$\mjj$                                                 & Dijet mass \\
$\detajj$                                              & Difference in pseudorapidity between the leading and subleading jets \\ 
$\dphijj$	         	                       & Difference in azimuth angles between the leading and subleading jets \\
$\pt^{\jet1}$		         	               & $\pt$ of the leading jet \\
$\pt^{\jet2}$		         	               & $\pt$ of the subleading jet \\
$\pt^{\ell_1}$                   	               & Leading lepton $\pt$ \\
$\pt^{\ell\ell}$                 	               & Dilepton $\pt$ \\
$z_{\ell_1}^{*}$	         	               & Zeppenfeld variable of the leading lepton \\
$z_{\ell_2}^{*}$	         	               & Zeppenfeld variable of the subleading lepton \\
$\ptmiss$                        	               & Missing transverse momentum \\
\hline
\end{tabular}
}

\end{table*}

\section{Systematic uncertainties}
\label{sec:systematics}

Several sources of systematic uncertainty in the cross section measurements can affect the rates and shapes of the distributions for the signal and background processes. For each source of uncertainty, the impact in different bins of the final distribution is considered as being fully correlated, whereas different sources of uncertainty are treated as uncorrelated.

The uncertainties in the integrated luminosity measurements for the data used in this analysis are 2.5, 2.3, and 2.5\% for the 2016, 2017, and 2018 data samples~\cite{CMS-PAS-LUM-17-001,CMS-PAS-LUM-17-004,CMS-PAS-LUM-18-002}, respectively. The total integrated luminosity has an uncertainty of 1.8\% because of the uncorrelated time evolution of some systematic effects.

The simulation of pileup events assumes a total inelastic $\Pp\Pp$ cross section of 69.2\unit{mb}, with an associated uncertainty of 5\%~\cite{ATLAS:2016pu,Sirunyan:2018nqx}. The impact of the pileup on the expected signal and background yields is less than 1\%.

Discrepancies in the lepton reconstruction and identification efficiencies between data and simulation are adjusted by applying corrections to all MC simulation samples. The efficiency corrections, which depend on the $\pt$ and $\eta$ of the lepton, are determined using $\PZ \to \ell\ell$ events in the $\PZ$ boson peak region~\cite{Sirunyan:2018fpa,Khachatryan:2015hwa}. The determination of the trigger efficiency leads to an uncertainty smaller than 1\% in the expected signal yield. The lepton momentum scale uncertainty is computed by varying the momenta of the leptons in the simulation by their uncertainties, and by repeating the analysis selection. The resulting uncertainties in the event yields are about 1\% for both electrons and muons. These uncertainties are 
treated as correlated across the three data sets.

The uncertainty in the calibration of the jet energy scale and resolution directly affects the selection efficiency of the jet multiplicity requirement and the $\ptmiss$ measurement. These effects are estimated by changing the jet energy in the simulation up- and downwards by one standard deviation. The uncertainty in the jet energy scale and resolution is 2--5\%, depending on the $\pt$ and $\eta$~\cite{Khachatryan:2016kdb}, and the impact on the expected signal and background yields is 1--4\%. 

Discrepancies in the \PQb\ tagging efficiency between data and simulation are adjusted by applying corrections to the simulated samples~\cite{Sirunyan:2017ezt}, which are estimated separately for correctly and incorrectly identified jets. Each set of values results in uncertainties in the \PQb\ tagging efficiency of about 
1--4\%, and the impact on the expected signal and background yields is about 1\%. The uncertainties in the jet energy scale and \PQb\ tagging are treated as uncorrelated across the three data sets.

The dominant theoretical uncertainties corresponding to the choice of the QCD renormalization and factorization scales are estimated by varying these scales independently up and down by a factor of two from their nominal values. The largest cross section variation, while excluding the two extreme variations where one scale is varied up and the other one down, is taken as the uncertainty . The PDF uncertainties are evaluated according to the procedure described in Ref.~\cite{PDFLHC}. The scale and PDF uncertainties for the processes estimated from simulation are treated as fully correlated across bins for the distributions used to extract the results. The effect of $\mathcal{O}(\alpha^7)$ correction for the unpolarized EW $\WW$ process on the shapes of the distributions for the $\WWLL$ and $\WWLT$ processes is considered as a systematic uncertainty. The correction values are used as a symmetric shape uncertainty. The uncertainties associated with the limited numbers of simulated events and of data events used to estimate the nonprompt lepton background are also included as systematic uncertainties with the latter being the dominant contribution. A summary of the systematic uncertainties in the $\WWLL$ and $\WWT$, and in the $\WWL$ and $\WWTT$ cross section measurements is shown in Table~\ref{tab:impactssswwlong_comb_exp}.

\begin{table*}[htb]
\centering
\topcaption{Systematic uncertainties of the $\WWLL$ and $\WWT$, and $\WWL$ and $\WWTT$ cross section 
measurements in units of percent.
\label{tab:impactssswwlong_comb_exp}}
\begin{tabular}{lcccc}
\hline
Source of uncertainty           & $\WWLL$ (\%) & $\WWT$ (\%) & $\WWL$ (\%) & $\WWTT$ (\%)  \\
\hline
Integrated luminosity           &  3.2  &  1.8  &  1.9  &  1.8  \\
Lepton measurement              &  3.6  &  1.9  &  2.5  &  1.8  \\
Jet energy scale and resolution & 11    &  2.9  &  2.5  &  1.1  \\
Pileup  		        &  0.9  &  0.1  &  1.0  &  0.3  \\
\PQb tagging   	        &  1.1  &  1.2  &  1.4  &  1.1  \\
Nonprompt lepton rate           & 17    &  2.7  &  9.3  &  1.6  \\
Trigger                         &  1.9  &  1.1  &  1.6  &  0.9  \\
Limited sample size             & 38    &  3.9  & 14    &  5.7  \\
Theory                          &  6.8  &  2.3  &  4.0  &  2.3  \\[\cmsTabSkip]

Total systematic uncertainty    & 44    &  6.6  & 18    &  7.0  \\[\cmsTabSkip]
Statistical uncertainty         & 123   & 15    & 42    & 22    \\[\cmsTabSkip]

Total uncertainty               & 130   & 16    & 46    & 23    \\
\hline
\end{tabular}
\end{table*}

\section{Results}
\label{sec:results}

Binned maximum-likelihood fits are performed to discriminate between the signals and the remaining backgrounds using the $\WW$ SR and the $\WZ$, nonprompt lepton, $\tZq$, and $\PZ\PZ$ CRs. Two separate fits are performed,
one for the simultaneous measurements of the $\WWLL$ and $\WWT$ cross sections and a second for the simultaneous measurements of the $\WWL$ and $\WWTT$ cross sections. The systematic uncertainties are treated as nuisance parameters and are profiled~\cite{Junk,Read1} with the shape and normalization of each distribution varying within the respective uncertainties in the fit. The normalization uncertainties are treated as log-normal nuisance parameters. The nuisance parameters are not significantly constrained. The small QCD $\WW$ contribution is normalized to the SM prediction and allowed to vary within the uncertainties. The normalizations of the $\tZq$, $\PZ\PZ$, and $\WZ$ background processes are free parameters of the maximum-likelihood fits, together with the signal cross sections. A two-dimensional distribution is used 
in the simultaneous fits for the $\WW$ SR with five bins in the inclusive BDT and five bins in the corresponding signal BDT. The $\mjj$ distribution is used for the CRs in the fit with four bins. The bin boundaries are chosen to have similar $\WWLL$ and $\WWL$ contributions across the bins.

The interference contributions between the EW and QCD diagrams for the $\WW$ and $\WZ$ processes are normalized to the SM predictions within the uncertainties. The impact of treatment of the interference contributions on the results is evaluated by performing a set of alternative fits where the interference contributions between the EW 
diagrams for the $\WWLL$ and $\WWT$ or $\WWL$ and $\WWTT$ processes and QCD diagrams are scaled with the square root of the measured to the predicted cross section ratios. The two approaches yield consistent results.  

\begin{figure*}[htbp]
\centering
\includegraphics[width=0.49\textwidth]{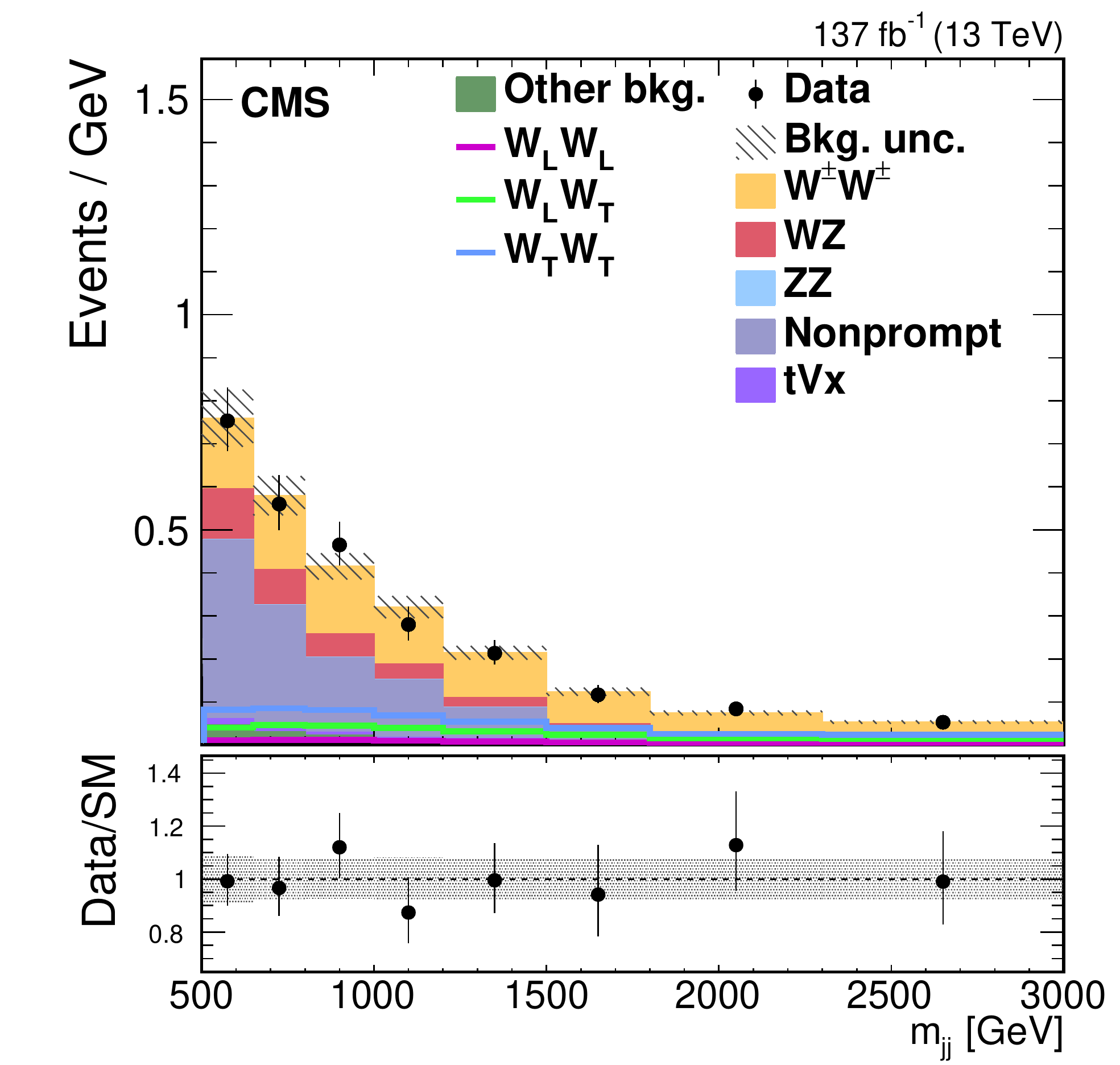}
\includegraphics[width=0.49\textwidth]{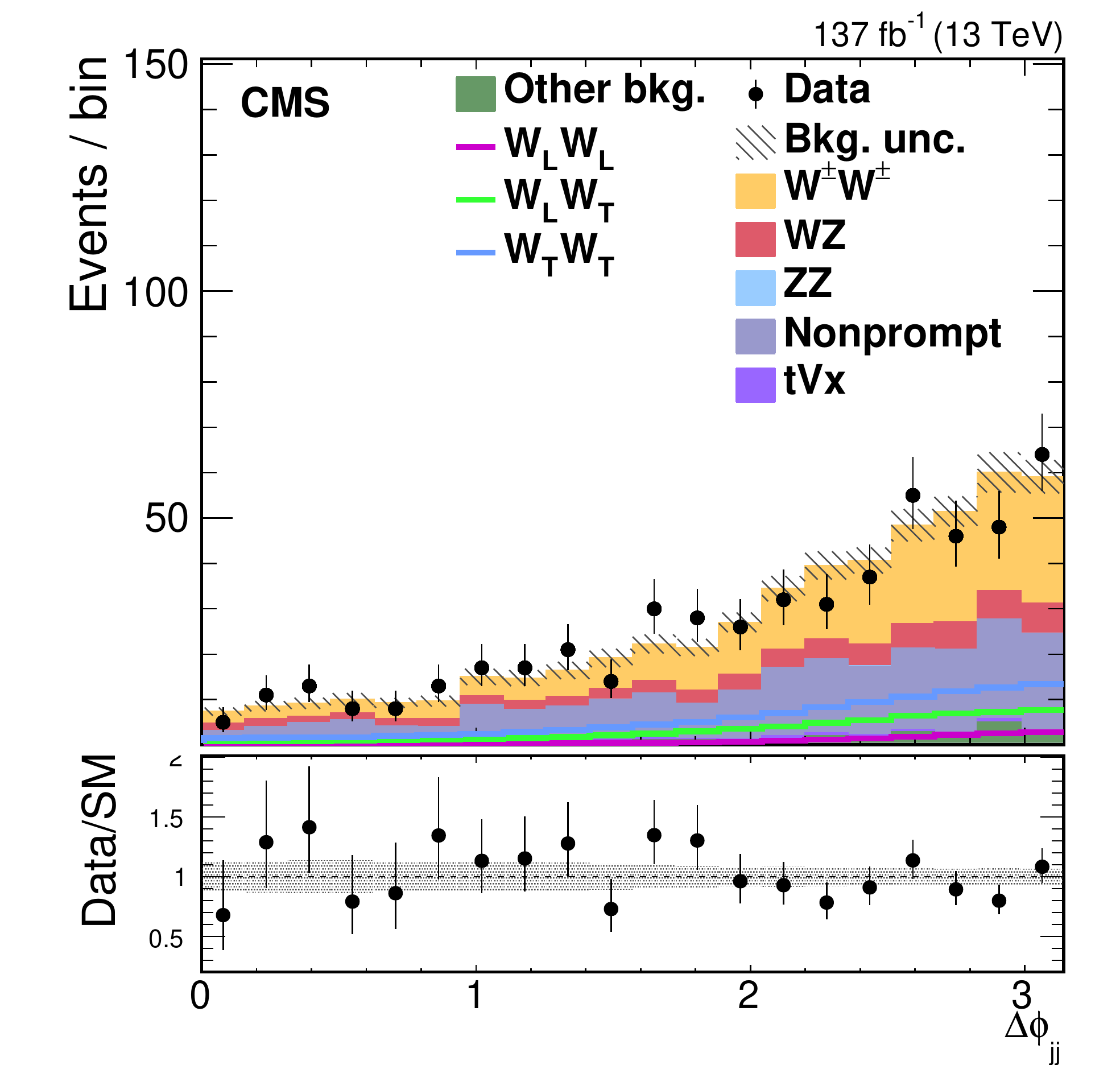}
\includegraphics[width=0.49\textwidth]{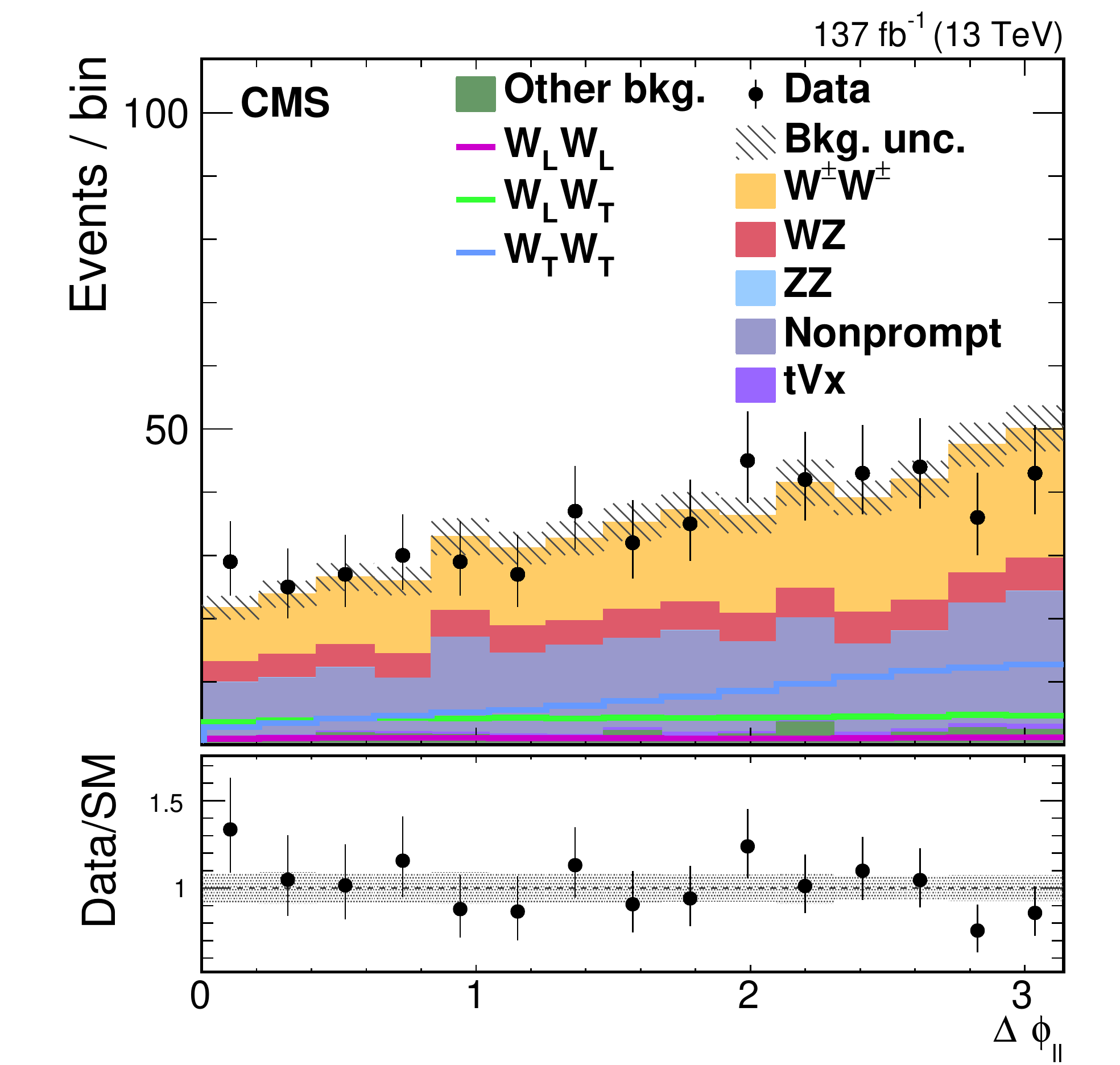}
\includegraphics[width=0.49\textwidth]{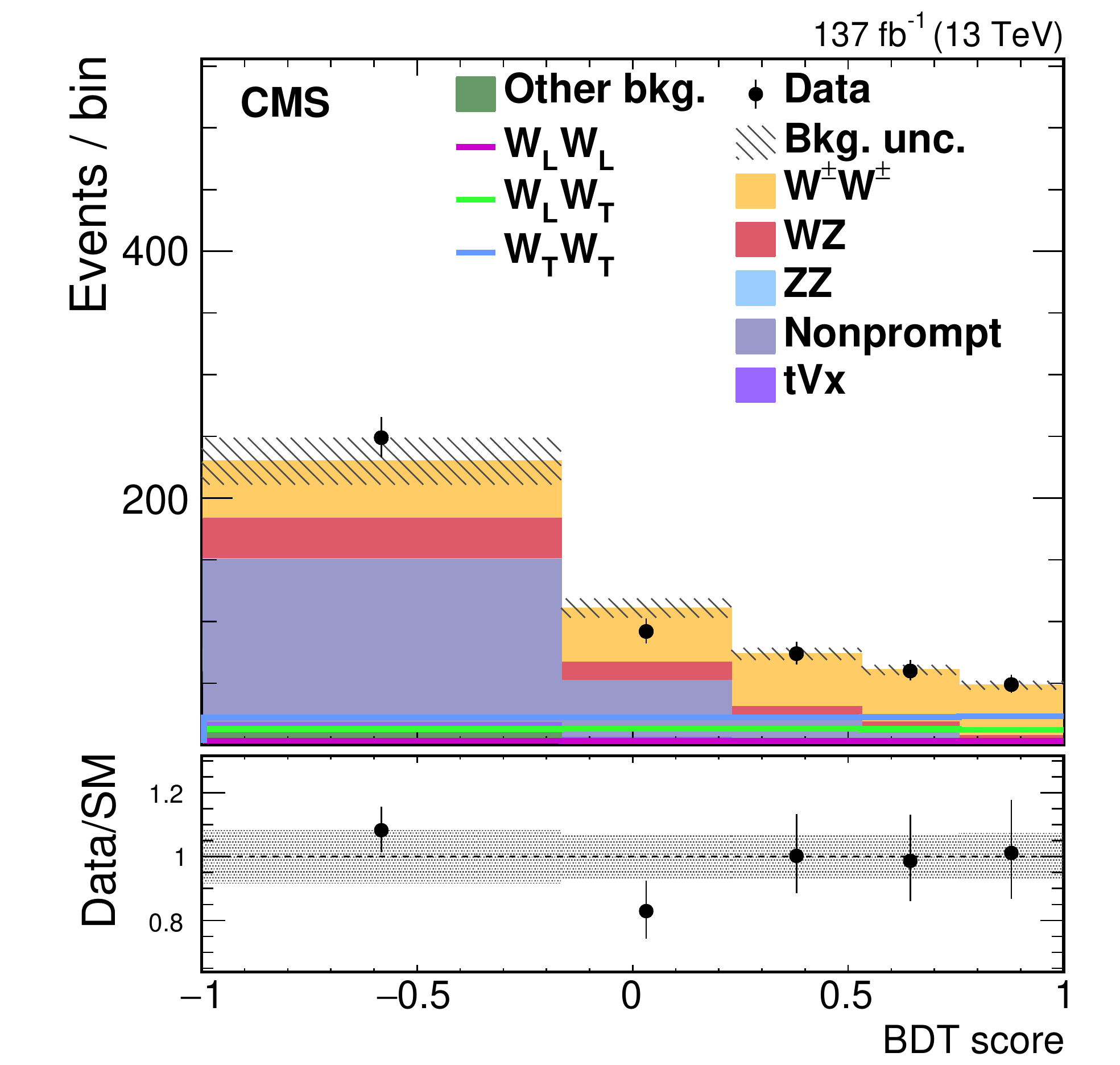}
\caption{
  Distributions of the $\mjj$ (upper left), $\dphijj$ (upper right), $\Delta\phi_{\ell\ell}$ (lower left), and of the output score of the inclusive BDT (lower right) in the $\WW$ SR. The predicted yields are shown with their best fit normalizations from the simultaneous fit. The histograms for the $\WW$ process include the contributions from the $\WWLL$, $\WWLT$, and  $\WWTT$ processes (shown separately as solid lines), QCD $\WW$, and interference. The histograms for other backgrounds  include the contributions from double parton scattering, $\PV\PV\PV$, and from oppositely charged dilepton final states from $\ttbar$, $\PQt\PW$, $\PW^{+}\PW^{-}$, and Drell--Yan processes. The overflow is included in the last bin. The bottom panel in each figure shows the ratio of the number of events observed in data to that of the total SM prediction. The gray bands represent the uncertainties in the predicted yields. The vertical bars represent the statistical uncertainties in the data.}
\label{fig:sswwlong_signalsel}
\end{figure*}

\begin{figure*}[htbp]
\centering
\includegraphics[width=0.49\textwidth]{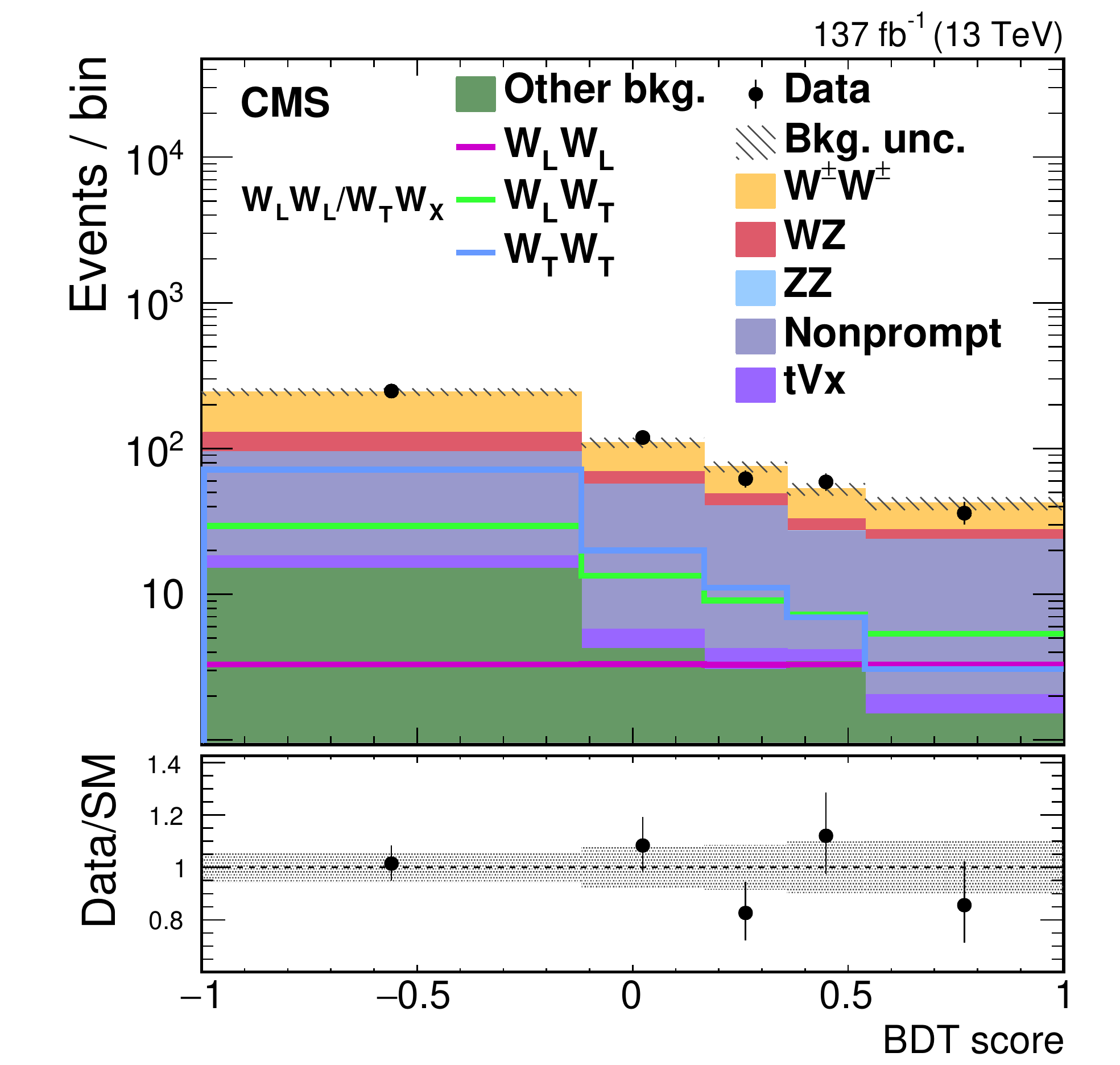}
\includegraphics[width=0.49\textwidth]{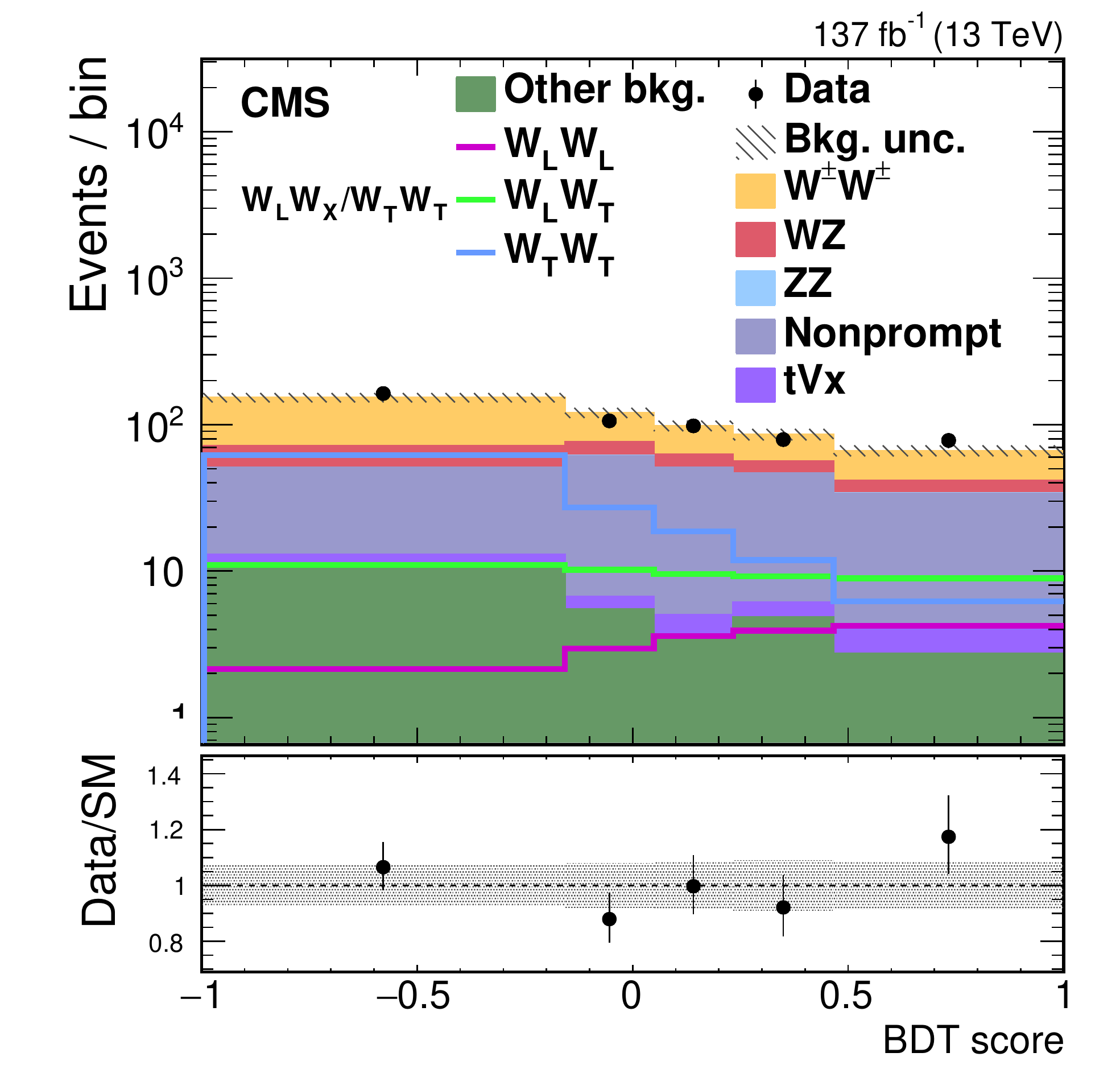}
\caption{
  Distributions of the output score of the signal BDT used for the $\WWLL$ and $\WWT$ cross section measurements (left) and of the output score of the signal BDT used for the $\WWL$ and $\WWTT$ cross section measurements (right). The predicted yields are shown with their best fit normalizations from the simultaneous fit. The histograms for the $\WW$ process include the contributions from the $\WWLL$, $\WWLT$, and $\WWTT$ processes (shown separately as solid lines), QCD $\WW$, and interference. The histograms for other backgrounds include the contributions from double parton scattering, $\PV\PV\PV$, and from oppositely charged dilepton final states from $\ttbar$, $\PQt\PW$, $\PW^{+}\PW^{-}$, and Drell--Yan processes. The bottom panel in each figure  shows the ratio of the number of events observed in data to that of the total SM prediction. The gray bands represent the uncertainties in the predicted yields. The vertical bars represent the statistical uncertainties in the data.}
\label{fig:sswwlong_signalsel2}
\end{figure*}

The distributions of $\mjj$ (upper left), $\dphijj$ (upper right), $\Delta\phi_{\ell\ell}$ (lower left), and the output score of the inclusive BDT (lower right) in the $\WW$ SR are shown in Fig.~\ref{fig:sswwlong_signalsel}. The distributions of the two signal BDT output scores are shown in Fig.~\ref{fig:sswwlong_signalsel2}. The predicted yields are shown with their best fit normalizations from the simultaneous fit for the $\WWLL$ and $\WWT$ cross sections. The data yields, together with the SM expectations with the best fit normalizations, are given in Table~\ref{tab:yield}. The background yields with the best fit normalizations from the simultaneous fit for the $\WWL$ and $\WWTT$ cross sections are consistent with the yields shown in Table~\ref{tab:yield} within a few percent.

\begin{table}[htb]
\centering
\topcaption{
  Expected yields from various SM processes and observed data events in $\WW$ SR. The combination of the statistical and systematic uncertainties is shown. The expected yields are shown with their best fit normalizations from the simultaneous fit for the $\WWLL$ and $\WWT$ cross sections. The $\WWLT$ and $\WWTT$ yields are obtained from the $\WWT$ yield assuming the SM prediction for the ratio of the yields. The $\tVx$ background yield includes the contributions from  $\ttbar\PV$ and $\tZq$ processes.\label{tab:yield}}
\newcolumntype{x}{D{,}{\,\pm\,}{3.3}}
\begin{tabular}{lx{c}@{\hspace*{5pt}}x}
\hline
Process &  \multicolumn{1}{c}{Yields in $\WW$ SR} \\
\hline
$\WWLL$                 &    16.0 ,  18.3 \\ 
$\WWLT$                 &    63.1 , 10.7 \\ 
$\WWTT$                 &   110.1 ,  18.1 \\ 
QCD $\WW$            &    13.8 ,   1.6 \\
Interference $\WW$   &     8.4 ,   0.6 \\
$\PW\PZ$               &    63.3 ,   7.8 \\
$\PZ\PZ$              &     0.7 ,  0.2 \\
Nonprompt               &   213.7 , 52.3 \\
$\tVx$                  &     7.1 ,  2.2 \\
Other background        &    26.9 ,  9.9 \\[\cmsTabSkip]

Total SM                &   522.9 ,  60.7 \\[\cmsTabSkip]
Data                    &  \multicolumn{1}{c}{524}       \\
\hline
\end{tabular}
\end{table}

The fiducial region for the cross section measurements is defined by requiring two same-sign leptons (electrons or muons) with $\pt>20\GeV$, $\abs{\eta}<2.5$, and $\mll>20\GeV$, and two jets with $\mjj>500\GeV$ and $\detajj>2.5$. The leptons at the generator level are selected at the so-called dressed level by combining the four-momentum of each lepton after final-state photon radiation with that of photons found within a cone of $\Delta R=0.1$ around the lepton. The jets at generator level are clustered from stable particles, excluding neutrinos, using the anti-$\kt$ clustering algorithm with a distance parameter of 0.4, and are required to satisfy $\pt>50\GeV$ and $\abs{\eta}<4.7$. Jets within $\Delta R<0.4$ of the selected charged leptons are not included. The overall signal selection efficiency within the fiducial region is about 40\%. Electrons and muons produced in the decay of a $\tau$ lepton are not included in the definition of the fiducial region. Nonfiducial signal events, $\ie$, events selected at the reconstructed level that do not satisfy the fiducial requirements, are scaled together with the fiducial signal events in the simultaneous fit. The relative contribution of the nonfiducial events is approximately 20\%. The nonfiducial events are treated as background processes.

\begin{figure}[htb]
\centering
\includegraphics[width=\cmsFigWidth]{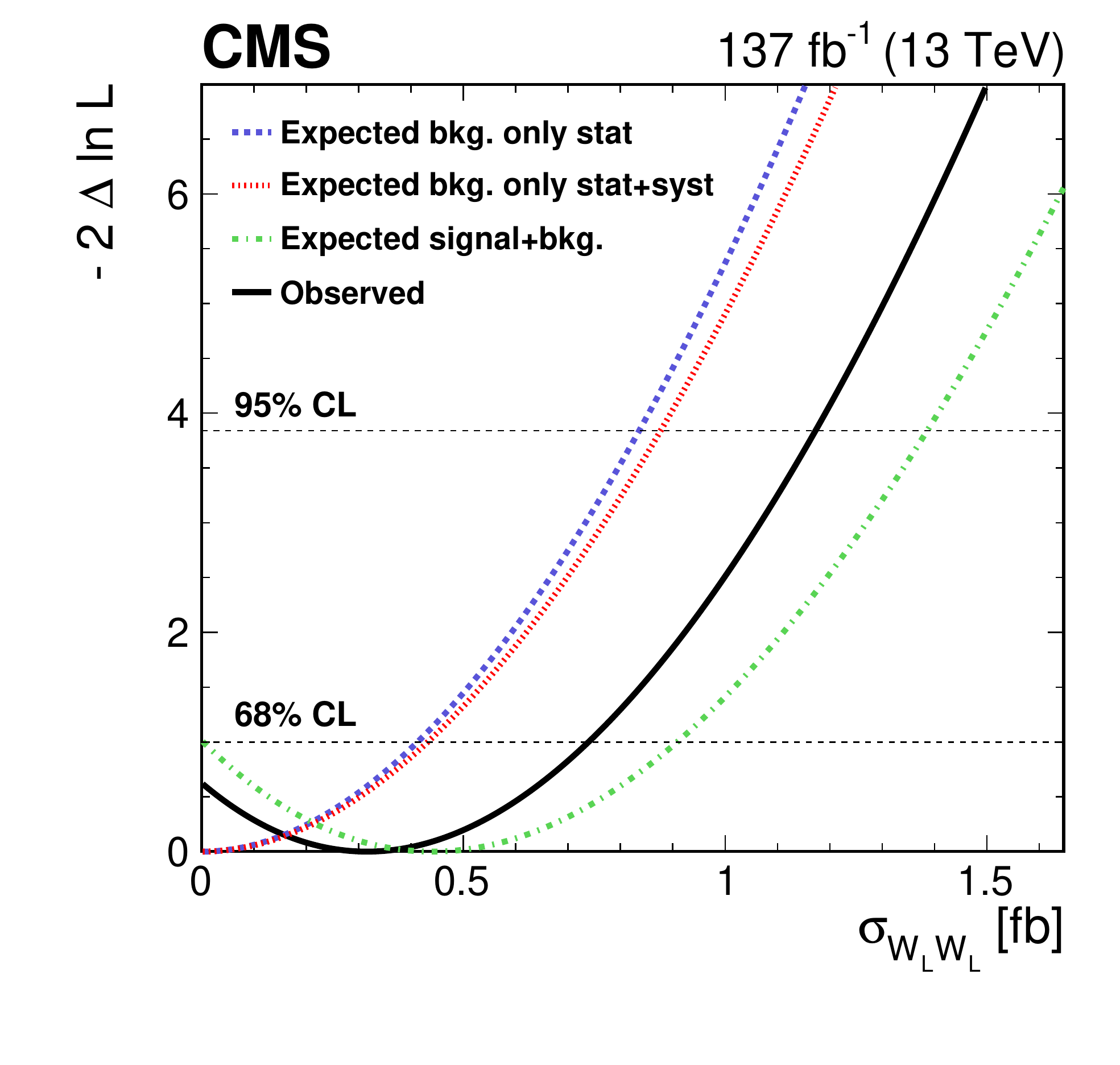}
\caption{
  Profile likelihood scan as a function of the $\WWLL$ cross section. The red (blue) line represents the expected values in the background-only hypothesis, \ie, assuming no contribution from the $\WWLL$ process, considering all systematic uncertainties (only statistical ones). The green line shows the expected values for the signal-plus-background hypothesis. The observed values are represented by the black line.
  \label{fig:scan}}

\end{figure}

The fit results  for the $\WWLL$ and $\WWT$ cross sections are shown in Fig.~\ref{fig:scan} as scans of the negative profile log-likelihood, $-2\Delta \text{ln} \mathcal{L}$, as a function of the $\WWLL$ cross section. The expected distributions include the contribution from the $\WWT$ process. The corresponding observed (expected) upper limit at 95\% confidence level (\CL) is 1.17 (0.88)\unit{fb}. The fiducial cross section measurements for the $\WWLL$ and $\WWT$ processes and the theoretical predictions are shown in Table~\ref{tab:default_sswwll}. The measured cross section values agree with the theoretical predictions within uncertainties.

The fiducial cross section measurements for the $\WWL$ and $\WWTT$ processes are extracted from a separate fit including the corresponding signal BDT. The measurements and the theoretical predictions are summarized in 
Table~\ref{tab:default_sswwll}. The significance of the measured $\WWL$ yield is quantified using background-only hypothesis, \ie, assuming no contribution from the $\WWL$ process, under the asymptotic approximation~\cite{CLs} and corresponds to 2.3 standard deviations. The expected significance is evaluated with an Asimov data set~\cite{CLs} and corresponds to 3.1 standard deviations.

\begin{table}[htb]
\topcaption{
  Measured fiducial cross sections for the $\WWLL$ and $\WWT$ processes, and for the $\WWL$ and $\WWTT$ processes for the helicity eigenstates defined in the $\WW$ center-of-mass frame. The combination of the statistical and systematic uncertainties is shown. The theoretical predictions including the  $\mathcal{O}(\alpS\alpha^6)$ and $\mathcal{O}(\alpha^7)$ corrections to the \MGvATNLO LO cross sections, as described in the text, are also shown. The theoretical uncertainties include statistical, PDF, and LO scale uncertainties; $\mathcal{B}$ is the branching fraction for $\PW\PW\to\ell\PGn\ell'\PGn$~\cite{10.1093/ptep/ptaa104}.
  \label{tab:default_sswwll}}
\centering
\begin{tabular}{ccc}
\hline
 Process                  & $\sigma \, \mathcal{B}$ (fb) & Theoretical prediction (fb) \\
\hline
$\WWLL$    & $0.32^{+0.42}_{-0.40}$  &  0.44 $\pm$ 0.05 \\
$\WWT$     & $3.06^{+0.51}_{-0.48}$  &  3.13 $\pm$ 0.35 \\[\cmsTabSkip] 

$\WWL$     & $1.20^{+0.56}_{-0.53}$  &  1.63 $\pm$ 0.18 \\
$\WWTT$    & $2.11^{+0.49}_{-0.47}$  &  1.94 $\pm$ 0.21 \\
\hline
\end{tabular}
\end{table}

The measurements are also performed for the polarized observables defined using the helicity eigenstates in the initial state parton-parton center-of-mass reference frame. Defining the polarization vectors in the parton-parton center-of-mass reference frame changes the respective contributions of $\WWLL$, $\WWL$ and $\WWT$, and the distributions of the input observables sensitive to the polarization~\cite{Ballestrero:2020qgv}. The fiducial cross section measurements and the theoretical predictions are summarized in Table~\ref{tab:default_sswwlx}. The observed (expected) 95\% \CL upper limit of the production cross section is 1.06 (0.85)\unit{fb} for the $\WWLL$ process. The observed (expected) significance of the $\WWL$ process is 2.6 (2.9) standard deviations.

\begin{table}[htb]
\topcaption{
  Measured fiducial cross sections for the $\WWLL$ and $\WWT$ processes, and for the $\WWL$ and $\WWTT$ processes for the helicity eigenstates defined in the parton-parton center-of-mass frame. The combination of the statistical and systematic uncertainties is shown. The theoretical predictions including the $\mathcal{O}(\alpS\alpha^6)$ and $\mathcal{O}(\alpha^7)$ corrections to the \MGvATNLO LO cross sections, as described in the text, are also shown. The theoretical uncertainties include statistical, PDF, and LO scale uncertainties; $\mathcal{B}$ is the branching fraction for $\PW\PW\to\ell\PGn\ell'\PGn$~\cite{10.1093/ptep/ptaa104}.
  \label{tab:default_sswwlx}}
\centering
\begin{tabular}{ccc}
\hline
 Process                  & $\sigma \, \mathcal{B}$ (fb) & Theoretical prediction (fb) \\
\hline
$\WWLL$    & $0.24^{+0.40}_{-0.37}$  &  0.28 $\pm$ 0.03 \\
$\WWT$     & $3.25^{+0.50}_{-0.48}$  &  3.32 $\pm$ 0.37 \\[\cmsTabSkip] 

$\WWL$     & $1.40^{+0.60}_{-0.57}$  &  1.71 $\pm$ 0.19 \\
$\WWTT$    & $2.03^{+0.51}_{-0.50}$  &  1.89 $\pm$ 0.21 \\
\hline
\end{tabular}
\end{table}

\section{Summary}
\label{sec:summary}

The first measurements of production cross sections for polarized same-sign $\WW$ boson pairs are reported. The measurements are based on a sample of proton-proton collisions at a center-of-mass energy of $13\TeV$ collected by the CMS detector at the LHC, corresponding to an integrated luminosity of $137\fbinv$. Events are selected by requiring exactly two same-sign leptons (electrons or muons), moderate missing transverse momentum, and two jets with a large rapidity separation and a high dijet mass. Boosted decision trees are used to separate between the polarized scattering processes by exploiting the kinematic differences. An observed (expected) 
95\% confidence level upper limit on the production cross section for longitudinally polarized same-sign $\WW$ boson pairs of 1.17 (0.88)\unit{fb} is reported with the helicity eigenstates defined in the $\WW$ center-of-mass reference frame. 
The electroweak production of the $\WW$ boson pairs where at least one of the $\PW$ bosons is longitudinally polarized is measured with an observed (expected) significance of 2.3 (3.1) standard deviations. Results are also 
reported with the polarizations defined in the parton-parton center-of-mass reference frame. The measured cross section values agree with the standard model predictions.

\begin{acknowledgments}
  We congratulate our colleagues in the CERN accelerator departments for the excellent performance of the LHC and thank the technical and administrative staffs at CERN and at other CMS institutes for their contributions to the success of the CMS effort. In addition, we gratefully acknowledge the computing centers and personnel of the Worldwide LHC Computing Grid for delivering so effectively the computing infrastructure essential to our analyses. Finally, we acknowledge the enduring support for the construction and operation of the LHC and the CMS detector provided by the following funding agencies: BMBWF and FWF (Austria); FNRS and FWO (Belgium); CNPq, CAPES, FAPERJ, FAPERGS, and FAPESP (Brazil); MES (Bulgaria); CERN; CAS, MoST, and NSFC (China); COLCIENCIAS (Colombia); MSES and CSF (Croatia); RIF (Cyprus); SENESCYT (Ecuador); MoER, ERC IUT, PUT and ERDF (Estonia); Academy of Finland, MEC, and HIP (Finland); CEA and CNRS/IN2P3 (France); BMBF, DFG, and HGF (Germany); GSRT (Greece); NKFIA (Hungary); DAE and DST (India); IPM (Iran); SFI (Ireland); INFN (Italy); MSIP and NRF (Republic of Korea); MES (Latvia); LAS (Lithuania); MOE and UM (Malaysia); BUAP, CINVESTAV, CONACYT, LNS, SEP, and UASLP-FAI (Mexico); MOS (Montenegro); MBIE (New Zealand); PAEC (Pakistan); MSHE and NSC (Poland); FCT (Portugal); JINR (Dubna); MON, RosAtom, RAS, RFBR, and NRC KI (Russia); MESTD (Serbia); SEIDI, CPAN, PCTI, and FEDER (Spain); MOSTR (Sri Lanka); Swiss Funding Agencies (Switzerland); MST (Taipei); ThEPCenter, IPST, STAR, and NSTDA (Thailand); TUBITAK and TAEK (Turkey); NASU (Ukraine); STFC (United Kingdom); DOE and NSF (USA).
   
  \hyphenation{Rachada-pisek} Individuals have received support from the Marie-Curie program and the European Research Council and Horizon 2020 Grant, contract Nos.\ 675440, 752730, and 765710 (European Union); the Leventis Foundation; the A.P.\ Sloan Foundation; the Alexander von Humboldt Foundation; the Belgian Federal Science Policy Office; the Fonds pour la Formation \`a la Recherche dans l'Industrie et dans l'Agriculture (FRIA-Belgium); the Agentschap voor Innovatie door Wetenschap en Technologie (IWT-Belgium); the F.R.S.-FNRS and FWO (Belgium) under the ``Excellence of Science -- EOS" -- be.h project n.\ 30820817; the Beijing Municipal Science \& Technology Commission, No. Z191100007219010; the Ministry of Education, Youth and Sports (MEYS) of the Czech Republic; the Deutsche Forschungsgemeinschaft (DFG) under Germany's Excellence Strategy -- EXC 2121 ``Quantum Universe" -- 390833306; the Lend\"ulet (``Momentum") Program and the J\'anos Bolyai Research Scholarship of the Hungarian Academy of Sciences, the New National Excellence Program \'UNKP, the NKFIA research grants 123842, 123959, 124845, 124850, 125105, 128713, 128786, and 129058 (Hungary); the Council of Science and Industrial Research, India; the HOMING PLUS program of the Foundation for Polish Science, cofinanced from European Union, Regional Development Fund, the Mobility Plus program of the Ministry of Science and Higher Education, the National Science Center (Poland), contracts Harmonia 2014/14/M/ST2/00428, Opus 2014/13/B/ST2/02543, 2014/15/B/ST2/03998, and 2015/19/B/ST2/02861, Sonata-bis 2012/07/E/ST2/01406; the National Priorities Research Program by Qatar National Research Fund; the Ministry of Science and Higher Education, project no. 02.a03.21.0005 (Russia); the Tomsk Polytechnic University Competitiveness Enhancement Program; the Programa Estatal de Fomento de la Investigaci{\'o}n Cient{\'i}fica y T{\'e}cnica de Excelencia Mar\'{\i}a de Maeztu, grant MDM-2015-0509 and the Programa Severo Ochoa del Principado de Asturias; the Thalis and Aristeia programs cofinanced by EU-ESF and the Greek NSRF; the Rachadapisek Sompot Fund for Postdoctoral Fellowship, Chulalongkorn University and the Chulalongkorn Academic into Its 2nd Century Project Advancement Project (Thailand); the Kavli Foundation; the Nvidia Corporation; the SuperMicro Corporation; the Welch Foundation, contract C-1845; and the Weston Havens Foundation (USA).
  
  \end{acknowledgments}

\bibliography{auto_generated}
\cleardoublepage \appendix\section{The CMS Collaboration \label{app:collab}}\begin{sloppypar}\hyphenpenalty=5000\widowpenalty=500\clubpenalty=5000\vskip\cmsinstskip
\textbf{Yerevan Physics Institute, Yerevan, Armenia}\\*[0pt]
A.M.~Sirunyan$^{\textrm{\dag}}$, A.~Tumasyan
\vskip\cmsinstskip
\textbf{Institut f\"{u}r Hochenergiephysik, Wien, Austria}\\*[0pt]
W.~Adam, T.~Bergauer, M.~Dragicevic, J.~Er\"{o}, A.~Escalante~Del~Valle, R.~Fr\"{u}hwirth\cmsAuthorMark{1}, M.~Jeitler\cmsAuthorMark{1}, N.~Krammer, L.~Lechner, D.~Liko, I.~Mikulec, F.M.~Pitters, N.~Rad, J.~Schieck\cmsAuthorMark{1}, R.~Sch\"{o}fbeck, M.~Spanring, S.~Templ, W.~Waltenberger, C.-E.~Wulz\cmsAuthorMark{1}, M.~Zarucki
% ...
\vskip\cmsinstskip
\dag: Deceased\\
1:  Also at Vienna University of Technology, Vienna, Austria\\
% ...
\end{sloppypar}
\end{document}